\documentclass[aps,pra,twocolumn,preprintnumbers,superscriptaddress,nofootinbib]{revtex4-1}
\usepackage{diagbox}
\usepackage{colortbl}
\usepackage{amsmath}
\usepackage{mathrsfs}
\usepackage{amsthm}
\usepackage{braket}
\usepackage{setspace}
\usepackage{titlesec}
\usepackage{amsfonts}
\usepackage{graphicx}
\usepackage{tikz}
\usepackage{pgfplots}
\usetikzlibrary{graphs}
\usepackage{color}
\usepackage{setspace}
\usepackage{bm}
\usepackage[shortlabels]{enumitem}
\usepackage{IEEEtrantools}
\usepackage{bbold}
\usepackage{mathtools}
\usepackage{amssymb}
\usepackage{dsfont}
\usepackage{color,soul}
\pgfplotsset{compat=1.10}
\usepgfplotslibrary{fillbetween}
\usetikzlibrary{patterns}
\usepackage{changes}

\makeatletter
\let\MYcaption\@makecaption
\makeatother

\usepackage[font=small]{subcaption}

\makeatletter
\let\@makecaption\MYcaption
\makeatother

\DeclareMathOperator*{\argmin}{argmin}

\newtheorem{thm}{Theorem}

\titleformat{\paragraph}[runin]{\itshape}{}{1em}{}[ --\hspace{5pt}  ]
\newcommand{\Q}{\mathcal{Q}}
\newcommand{\I}{\mathcal{I}}
\newcommand{\Hmin}{H_{\text{min}}}

\newcommand{\Nraw}{N_{\text{raw}}}
\newcommand{\Nest}{N_{\text{est}}}
\newcommand{\Ntot}{N_{\text{tot}}}

\newcommand{\Preg}{P^{\text{reg}}_{AB|XY}}
\newcommand{\PXYchi}{P_{XY}^{\chi}}

\definecolor{nred}{rgb}{1,0,0}
\definecolor{nblack}{rgb}{0,0,0}
\definecolor{nblue}{rgb}{0.2,0.2,0.8}
\definecolor{ngreen}{rgb}{0.2,0.6,0.2}

\usepackage{hyperref}

\begin{document}

\author{Boris Bourdoncle}
\email{boris.bourdoncle@icfo.eu}
\affiliation{ICFO-Institut de Ci\`encies Fot\`oniques, The Barcelona Institute of Science and Technology, 08860 Castelldefels (Barcelona), Spain}
\author{Pei-Sheng Lin}
\email{pslin@phys.ncku.edu.tw}
\affiliation{Department of Physics, National Cheng Kung University, Tainan 701, Taiwan}
\author{Denis Rosset}
\affiliation{Department of Physics, National Cheng Kung University, Tainan 701, Taiwan}
\affiliation{Perimeter Institute for Theoretical Physics, Waterloo, Ontario Canada N2L 2Y5}
\author{Antonio Ac\'in}
\affiliation{ICFO-Institut de Ci\`encies Fot\`oniques, The Barcelona Institute of Science and Technology, 08860 Castelldefels (Barcelona), Spain}
\affiliation{ICREA-Instituci\'o Catalana de Recerca i Estudis Avan\c cats, Lluis Companys 23, 08010 Barcelona, Spain}
\author{Yeong-Cherng Liang}
\affiliation{Department of Physics, National Cheng Kung University, Tainan 701, Taiwan}

\title{Regularising data for practical randomness generation}

\begin{abstract} 
Assuming that the no-signalling principle holds, non-local correlations contain intrinsic randomness. In particular, for a specific Bell experiment, one can derive relations between the amount of randomness produced, as quantified by the min-entropy of the output data, and its associated violation of a Bell inequality. In practice, due to finite sampling, certifying randomness requires the development of statistical tools to lower-bound the min-entropy of the data as a function of the estimated Bell violation. The quality of such bounds relies on the choice of certificate, i.e., the Bell inequality whose violation is estimated. In this work, we propose a method for choosing efficiently such a certificate and analyse, by means of extensive numerical simulations (with various choices of parameters), the extent to which it works. The method requires sacrificing a part of the output data in order to estimate the underlying correlations. Regularising this estimate then allows one to find a Bell inequality that is well suited for certifying practical randomness from these specific correlations. We then study the effects of various parameters on the obtained min-entropy bound and explain how to tune them in a favourable way. Lastly, we carry out several numerical simulations of a Bell experiment to show the efficiency of our method: we nearly always obtain higher min-entropy rates than when we use a pre-established Bell inequality, namely the Clauser-Horne-Shimony-Holt inequality. 
\end{abstract}

\maketitle

\section{Introduction}

Being able to produce bits that are impossible to predict is crucial for a number of cryptographic tasks. In order to characterise the unpredictability of the outcomes of a given experiment, one usually models an adversary who has access to some information on the devices used in the experiment.  Bounds on how well the adversary can predict the output bits, conditioned on the information this adversary was given, can then be derived. If the devices in use behave classically, and if the adversary is given total information about them, no unpredictable bits can be obtained, as classical physics is deterministic. By contrast, if the devices are quantum, their outputs can be impossible to predict, even when the adversary has access to a perfect characterisation of the devices. 
 
In practice, a perfect control of quantum devices is rarely possible. This means that, in most cases, even the users do not have access to a perfect characterisation of the devices. Fortunately, the unpredictability of a sequence of bits can be certified even when the devices producing them cannot be completely characterised, thanks to the device-independent approach to quantum information protocols \cite{Colbeck2006Quantum,Mayers1998Quantum,Acin2007Device, Pironio2010Random, Acin2016Certified}. In this case, the minimal requirement is two separated devices that each receives an input --- measurement choice --- and produces an output --- measurement result --- without communicating. This is usually called a Bell experiment. The key idea is as follows: if the input-outputs correlations are `Bell non-local'~\cite{Brunner2014Bell} (hereafter abbreviated `non-local'), the outputs cannot be deterministic, irrespective of the extent to which the devices can be, or have been characterised. That is to say, they contain some intrinsic randomness.  

Quantifying the unpredictability of the bits obtained in a Bell experiment is not a trivial task, as it depends on a number of factors, including how powerful the adversary is assumed to be \cite{Vazirani2012Certifiable}, how the devices are assumed to behave with time \cite{Arnon2012Limits, Ramanathan2016Randomness} or how the users process the accessible information \cite{Bancal2014More, Nieto2018Device}. 

In this work, we adopt the most common approach to estimating the unpredictability of a Bell experiment: a user enters a bit in each of two shielded devices, which in return give output bits, according to some conditional probability distribution. These bits can be used to compute the violation of a Bell inequality --- a constraint necessarily satisfied by physical devices that function in a local deterministic manner. Given this violation, an eavesdropper designs an optimal strategy for guessing the output bits. Here, we restrict our attention to the case where the adversary obeys quantum mechanics, i.e., we do not study the case of a supra-quantum eavesdropper~\cite{Pironio2010Random}. Moreover, we assume that the eavesdropper only has access to classical side information~\cite{Fehr2013Security, Knill2017Rand, Nieto2018Device} (for the case of an adversary with  quantum side information, we refer the readers to \cite{Dupuis2016Entropy, Knill2018QuantumAdversary, Arnon2018Practical}). In the honest provider scenario that we consider here, that is the appropriate level of security, since device-independent randomness generation involves only one user in one location: no quantum information needs to be sent over an insecure channel. The only thing the adversary may exploit in this case is the imperfection of the device such as noise or deterioration with time. We refer the readers to \cite{Pironio2013SecurityRandomness} for a detailed explanation. We then quantify the randomness of the sequence of output bits by its min-entropy. 

The upside of this approach is its simplicity, as it depends on only one parameter: the violation of a Bell inequality. However, in a real Bell experiment, this number cannot be exactly known, as the number of runs is finite. One can only compute an estimate of the average Bell violation. To overcome this obstacle, statistical tools were developed that allow one to upper-bound the predictability of the outputs with arbitrary confidence, based only on an estimate of the Bell violation, rather than its theoretical value \cite{Pironio2010Random, Fehr2013Security, Pironio2013SecurityRandomness, Nieto2018Device}. See also \cite{Knill2017Rand} for other type of statistical tool. 

Another question naturally arises in this approach: which Bell inequality should one use to obtain good bounds? A Bell inequality violation contains only partial information about the input-output correlation. Choosing the inequality poorly can result in a serious underestimation of the unpredictability of a Bell experiment, and may not even certify any unpredictability, as every non-local correlations satisfy some Bell inequalities. Yet, if the input-output distribution is known, finding the Bell inequality that certifies as much randomness as possible turns out to be a semi-definite program \cite{Nieto2014Using, Bancal2014More}. Unfortunately, as mentioned above, the input-output distribution is not accessible in practice, due to finite statistics. 

We thus propose a method to circumvent this problem. It consists in using part of the input-output statistics to estimate the corresponding underlying distribution. It is however very likely that a naive estimate based on the relative frequencies will not correspond to a distribution achievable with quantum physics. Consequently, the above-mentioned semi-definite program is not directly applicable as it can only be solved for distribution that belongs to the quantum set, or to some specific relaxation of this set, defined by the Navascu\'es-Pironio-Ac\'in (NPA) hierarchy \cite{Navascues2007Bounding, Navascues2008Convergent}. We thus employ the methods developed in \cite{Lin2018Device} in order to obtain a distribution approximating the underlying distribution that lies inside one the NPA sets. This then enables us to solve the corresponding semi-definite program and hence obtain a Bell inequality specifically suited for the estimated distribution, and hence better tailored for the underlying distribution. 

The rest of this article is organised as follows. In Section \ref{sec:Theory}, we remind some known results about how to lower bound the min-entropy associated to a practical Bell experiment. In Section \ref{sec:Results}, we present our results. It consists of a method to optimise the choice of Bell inequality in order to improve the bound on the min-entropy. We then study the effects of various parameters of our method on a few behaviours picked at random in order to tune them favourably. Finally, we  demonstrate the efficiency of this method by presenting our numerical results obtained by running various numerical simulations of Bell experiments. We conclude with some open questions and possible future works in Section \ref{sec:Conclusion}.   

\section{Lower bound on the min-entropy}
\label{sec:Theory}

We now remind the framework commonly used to quantify the randomness generated in a theoretical Bell test, and the mathematical tools developed to lower bound the randomness generated in a real Bell experiment. Here, by a theoretical Bell test, we mean the ideal, asymptotic situation where the underlying distribution is attained. By contrast, in a real experiment, the data available is subjected to statistical fluctuations.

\subsection{Preliminaries} 

We define a Bell test in the following way: a user has access to two devices $\mathcal{A}$ and $\mathcal{B}$. The internal working of those devices is unknown: they are treated as black boxes. The only possible interaction with those black boxes is as follows: upon receiving an input $x \in \{0,1\}$ (resp. $y \in \{0,1\}$), $\mathcal{A}$ (resp. $\mathcal{B}$) produces an output $a \in \{0,1\}$ (resp. $b \in \{0,1\}$). We associate the random variables $A,B,X$ and $Y$ to $a,b,x$ and $y$ respectively, and $P_{AB|XY}$ denotes the conditional probabilities of the outputs given the inputs, which we call hereafter a behaviour (the subscript indicating the random variables is sometimes omitted when they are clear from the context). We assume that this behaviour obeys quantum mechanics, i.e., $P_{AB|XY}(ab|xy) = \text{Tr}[\rho\, M^A_{a|x}\otimes M^B_{b|y}]$, where $\rho$ is a quantum state and $\{M^\mathrm{A}_{a|x}\}_a$ and $\{M^\mathrm{B}_{b|y}\}_b$ are positive-operator valued measures. This implies in particular that input $x$ (resp. $y$) has no influence on output $b$ (resp. $a$), i.e., $P_{AB|XY}$ is no-signalling~\cite{Popescu1994Quantum,Barrett2005Nonlocal}. We denote by $\Q$ the set of all quantum behaviours.

When a Bell test is repeated $n$ times, we write $\mathbf{x}=(x_1,...,x_n)$ for the sequence of inputs of $\mathcal{A}$. We define $\mathbf{y}, \mathbf{a}, \mathbf{b}$, as well as their associated random variables $\mathbf{A}, \mathbf{B}, \mathbf{X}, \mathbf{Y}$, in the same way. We now briefly remind some of the key concepts that we will use later on.  

\paragraph*{Min-entropy} 
We quantify the randomness of the outputs produced in a Bell test via the min-entropy. The min-entropy of $(\mathbf{A},\mathbf{B})$ given $(\mathbf{X},\mathbf{Y})$ conditioned on some event $\lambda$, according to a distribution $P=P_{\mathbf{A}\mathbf{B}\mathbf{X}\mathbf{Y}}$, is:
\begin{multline}
\Hmin(\mathbf{A},\mathbf{B}|\mathbf{X},\mathbf{Y}, \lambda)_{P} = \\
-\log_2 \sum_{\mathbf{x},\mathbf{y}}P(\mathbf{x},\mathbf{y}|\lambda) \max_{\mathbf{a},\mathbf{b}}P(\mathbf{a},\mathbf{b}|\mathbf{x},\mathbf{y},\lambda).
\end{multline}
Essentially, the min-entropy quantifies the number of almost-uniform random bits that can be obtained from a source via a randomness extractor. The event $\lambda$ is typically a function of the specific inputs that were chosen and the specific outputs that were obtained during the Bell experiment, such as a statistical estimate. For a detailed review on the relevance of this quantity, see \cite{Konig2009Operational}. 

We now introduce all the elements that allow us to lower-bound this quantity. 

\paragraph*{Bell expression}
We call a real linear functional in $P_{AB|XY}$ a Bell expression:
\begin{equation}
	\mathcal{I}(P_{AB|XY})=\sum_{a,b,x,y}c_{abxy}P_{AB|XY}(ab|xy).
\label{BellExpression}
\end{equation}
For a given Bell expression, its maximal value over all local deterministic strategies, i.e., all behaviours with $a$ (resp. $b)$ being a deterministic function of $x$ (resp. $y)$, gives rise to the local bound $I_{\mathcal{L}}$.      
A behaviour is said to be local if it can be written as a convex mixture of deterministic strategies, and the corresponding set is denoted $\mathcal{L}$. The inequality:
\begin{equation}
\forall P_{AB|XY} \in \mathcal{L}, \quad \mathcal{I}(P_{AB|XY}) \leq I_{\mathcal{L}}
\end{equation}
is referred to as a Bell inequality \cite{Bell1964}.

However, in quantum physics, non-local behaviours are accessible. For a given $P_{AB|XY}$, it might then happen that this local bound is violated. In this case, we call Bell violation the value that the Bell expression takes, and we denote by $I_{\mathcal{Q}}^+$ the maximal value allowed in quantum theory, and  by $I_{\mathcal{Q}}^-$, respectively, the minimal value: 
\begin{equation}
I_{\mathcal{Q}}^+=\max_{P \in \mathcal{Q}} \mathcal{I}(P),\quad I_{\mathcal{Q}}^-=\min_{P \in \mathcal{Q}} \mathcal{I}(P).
\end{equation}
As is usually done, we call `local bound' and `quantum bound' the maximal values of a Bell expression over the local and quantum sets. However, we also introduce the minimal value of a Bell expression over the quantum set, as we need it to express the min-entropy bound given by Theorem \ref{Theorem} (see \cite{Pironio2013SecurityRandomness, Nieto2018Device} for details).  

\paragraph*{Observed frequencies}
For a given realisation of $(\mathbf{A},\mathbf{B},\mathbf{X},\mathbf{Y})$, we define the observed frequencies as: 
\begin{equation}
\hat{P}_{AB|XY}(ab|xy) = \frac{N_{abxy}}{N_{xy}},
\end{equation} 
where $N_{abxy}$ (resp. $N_{xy}$) is the number of occurrences of the quadruplet $(a,b,x,y)$ (resp. the pair $(x,y)$) in the $n$ length sequence $(\mathbf{a},\mathbf{b}, \mathbf{x}, \mathbf{y})$. In the unlikely event that a given pair $(x^*,y^*)$ was never input, i.e., $N_{x^*y^*}=0$, the experiment should be performed again.

\paragraph*{Observed Bell violation}
For simplicity, we assume that the inputs $(\mathbf{x},\mathbf{y})$ are chosen independently and identically at each round with probability $P(X_i=x,Y_i=y)=\pi_{xy}$. For a given Bell expression, as defined in equation \eqref{BellExpression}, and a given realisation of $(\mathbf{A},\mathbf{B},\mathbf{X},\mathbf{Y})$, we define the observed average Bell violation as: 
\begin{equation}
\hat{I}= \sum_{a,b,x,y} c_{abxy} \frac{N_{abxy}}{n\cdot\pi_{xy}}.
\end{equation}  

We point out that, even though $\hat{P}$ and $\hat{I}$ are both estimators, they do not involve the inputs in the same manner. To compute $\hat{P}$, one counts the occurrences of both the quadruplets $(a,b,x,y)$ and the input pairs $(x,y)$, whereas for $\hat{I}$, one only counts the quadruplets $(a,b,x,y)$ and uses directly the input distributions $\pi_{xy}$, instead of the frequencies of each input pair for a given realisation. Both can be computed from a realisation of Bell experiments, as $\pi_{xy}$ is chosen by the user (see details hereafter). However, we decide to compute the observed frequencies using $N_{xy}$ to ensure that $\hat{P}_{AB|X=x,Y=y}$ is normalised for each $(x,y)$, and can thus be identified as a probability distribution. On the other hand, we decide to compute the observed Bell violation $\hat{I}$ directly using the input distribution, as this is crucial for the derivation of Theorem \ref{Theorem} (see \cite{Pironio2013SecurityRandomness, Nieto2018Device} for details).  
Note that, if the behaviours of the devices at each round are independent and identically distributed (i.i.d.) according to some distribution $P_{AB|XY}$, $\hat{I}$ converges towards $\mathcal{I}(P_{AB|XY})$ when $n$ tends to infinity. However, we do not need to make such an assumption to define this quantity.
\paragraph*{Distance between distributions}
We say that two distributions $P_{\mathbf{A}\mathbf{B}\mathbf{X}\mathbf{Y}}$ and $\tilde{P}_{\mathbf{A}\mathbf{B}\mathbf{X}\mathbf{Y}}$ are $\epsilon$-close if their total variation distance is upper bounded by $\epsilon$:
\begin{equation}
d(P,\tilde{P}) = \frac{1}{2} \sum_{\mathbf{a},\mathbf{b},\mathbf{x},\mathbf{y}} | P(\mathbf{a},\mathbf{b},\mathbf{x},\mathbf{y}) - \tilde{P}(\mathbf{a},\mathbf{b},\mathbf{x},\mathbf{y})| \leq \epsilon.
\end{equation}
 
\paragraph*{Randomness-bounding function}
For a given Bell expression $\mathcal{I}$, let $\mathcal{I}(\mathcal{Q})=\{ \mathcal{I}(P) | P \in \mathcal{Q} \} $. Let $\chi$ be a subset of $\{0,1\}^2$. We say that $ H_{\mathcal{I}}^{\chi} :  \mathcal{I}(\mathcal{Q}) \to [0,2]$ is a randomness-bounding function (RB function) for $\chi$ if the two following requirements are satisfied:
\begin{enumerate}[label={R.\arabic*}]
\item $ \forall P \in \mathcal{Q}, \ \min\limits_{\substack{(a,b) \in \{0,1\}^2 \\ (x,y) \in \chi}} ( -\log_2 P(ab | xy) ) \geq H_{\mathcal{I}}^{\chi} (\mathcal{I}(P)), $ \label{Req1}
\item $H_{\mathcal{I}}^{\chi} $ is convex. \label{Req2}
\end{enumerate}
These requirements are needed in order to bound the min-entropy produced by a sequence of Bell tests (see \cite{Nieto2018Device} for a detailed explanation). $\chi$ specifies a subset of all possible inputs for which the RB function is valid. It should contain the inputs for which the associated conditional distributions are the most random, i.e., the inputs that yield  the largest $H_{\mathcal{I}}^{\chi} $. For instance, if one obtains a high $H_{\mathcal{I}}^{\chi} $ from one pair of input $(x^*,y^*)$, and a small $H_{\mathcal{I}}^{\chi} $ for the others, one would have an interest in setting $\chi$ to $(x^*,y^*)$ only. Indeed, the space over which the minimisation is carried out gets bigger when one includes more input pairs in $\chi$, which results in a smaller RB function, which, in turn, will give a smaller lower bound on the min-entropy. The reason for that will become clear in the next section. However, this trade-off depends on the total number of Bell tests that are used for generating randomness, as is illustrated by the numerical simulations presented hereafter.  

\subsection{The guessing probability}

The main ingredient needed to lower-bound the min-entropy is the RB function. We now explain how to compute it via the guessing probability problem. This general form was introduced and extensively explained in \cite{Nieto2018Device}. Here, we only briefly present the reasoning that leads to this formulation. For a given Bell expression $\mathcal{I}$ and a specific  value $I^*$ of $\I$, finding the lower bound $H_{\mathcal{I}}^{\chi}(I^*)$ defined by requirements \ref{Req1} and \ref{Req2} amounts to solving a minimisation problem over all quantum behaviours $P$ such that $\mathcal{I}(P)=I^*$. However, the optimisation problem obtained in this way is not easily solvable, due to the presence of the logarithm and to the complicated nature of the quantum set $\mathcal{Q}$~\cite{Navascues2007Bounding, Navascues2008Convergent,Goh2017Geometry}. 

This led the authors of \cite{Nieto2018Device} to consider instead the following problem. For $(\alpha,\beta) \in \{0,1\}^2$ and $(\gamma,\delta) \in \chi \subseteq \{0,1\}^2$, let $\{ \tilde{P}^{\alpha \beta \gamma \delta} \}$ be $4 \times |\chi|$ variables, where $|\chi|$ is the cardinality of the set $\chi$, that represent unnormalised behaviours. The problem then reads:
\begin{equation}
\begin{array}{>{\displaystyle}r<{\displaystyle}>{\displaystyle}c<{\displaystyle}>{\displaystyle}l<{\displaystyle}}
G^{\chi}_{\mathcal{I}}(I^*)  = & \max\limits_{\{ \tilde{P}^{\alpha \beta \gamma \delta}\}}  & \sum_{\substack{\alpha, \beta \in \{0,1 \}^2 \\ \gamma, \delta \in \chi}} \tilde{P}^{\alpha \beta \gamma \delta}(\alpha \beta | \gamma \delta) \\
& \mathrm{s.t.} & \sum_{\substack{\alpha, \beta \in \{0,1 \}^2 \\ \gamma, \delta \in \chi}} \mathcal{I}(\tilde{P}^{\alpha \beta \gamma \delta})= I^*, \\
& & \sum_{\substack{\alpha, \beta \in \{0,1 \}^2 \\ \gamma, \delta \in \chi}} \mathrm{Tr} [\tilde{P}^{\alpha \beta \gamma \delta}] = 1, \\
&  & \forall \ \alpha, \beta, \gamma, \delta, \ \tilde{P}^{\alpha \beta \gamma \delta} \in \mathcal{\tilde{Q}}_k,
\end{array}
\label{GI}
\end{equation}
where $\mathrm{Tr} [\tilde{P}] = \sum_{ab} \tilde{P}(ab|xy)$ is the norm of $\tilde{P}$ (which is independent of $(x,y)$ by no-signalling) and  $\tilde{\mathcal{Q}}_k$ is the set of unnormalised behaviours that belong to the $k^{th}$ level of the NPA hierarchy \cite{Navascues2007Bounding, Navascues2008Convergent}. This problem is then a semi-definite program (SDP), and, as such, can be efficiently solved. Moreover, if we let $H_{\mathcal{I}}^{\chi} =-\log_2 G^{\chi}_{\mathcal{I}}$, $H_{\mathcal{I}}^{\chi} $ satisfies both requirements \ref{Req1} and \ref{Req2}, and is thus a RB function for $\chi$ (see \cite{Nieto2018Device} for details). It is, however, not necessarily tight, in particular because the NPA hierarchy is merely a relaxation of $\Q$.

In the case where $\chi$ contains only one input pair, the guessing probability problem has a simple interpretation: it is the maximal guessing probability, over all quantum strategies, of an adversary who is bound to keep the Bell violation $I^*$ unchanged. This problem was introduced in \cite{Bancal2014More, Nieto2014Using}, along with another optimisation problem that we now remind. The idea is the following: if we consider the guessing probability as a theoretical measure of the randomness of a behaviour $P_{AB|XY}$, constraining this behaviour to only a Bell violation, that is, constraining only a linear functional of $P_{AB|XY}$ to a fixed value, amounts to discarding some information about the behaviour. It might thus result in an underestimation of the intrinsic randomness contained in $P_{AB|XY}$. On the contrary, the following problem takes into account the complete information about the behaviour to evaluate its randomness:
\begin{equation}
\begin{array}{>{\displaystyle}r<{\displaystyle}>{\displaystyle}c<{\displaystyle}>{\displaystyle}l<{\displaystyle}}
G^{\chi}_{full}(P)  = & \max\limits_{\{ \tilde{P}^{\alpha \beta \gamma \delta}\}}  &\sum_{\substack{\alpha, \beta \in \{0,1 \}^2 \\ \gamma, \delta \in \chi}} \tilde{P}^{\alpha \beta \gamma \delta}(\alpha \beta | \gamma \delta) \\
& \mathrm{s.t.} &\sum_{\substack{\alpha, \beta \in \{0,1 \}^2 \\ \gamma, \delta \in \chi}} \tilde{P}^{\alpha \beta \gamma \delta}= P, \\
&  & \forall \ \alpha, \beta, \gamma, \delta, \ {\tilde{P}}^{\alpha \beta \gamma \delta}\in \mathcal{\tilde{Q}}_k.
\end{array}
\label{GFull}
\end{equation}

As problem \eqref{GFull} is more constrained than problem \eqref{GI}, it is clear that $G^{\chi}_{full}(P) \leq G^{\chi}_{\mathcal{I}}(\mathcal{I}(P))$. One could compare these two problems in the following way: $G^{\chi}_{full}(P)$ is a measure of the randomness of a behaviour $P$, whereas $G^{\chi}_{\mathcal{I}}(I^*)$ is measure of the randomness that can be certified by a Bell expression $\mathcal{I}$. Yet these two formulations are connected: the dual problem of \eqref{GFull} precisely returns a Bell expression $\mathcal{I}^*$ such that $G_{\mathcal{I}^*}(\mathcal{I}^*(P))=G_{full}(P)$ \cite{Bancal2014More,Nieto2014Using}. When the Bell expression is well chosen, \eqref{GI} and \eqref{GFull} are thus equivalent. 

Let us stress however that these quantities can only be considered as theoretical measures of randomness for theoretical objects such as probability distributions and Bell expressions. In order to obtain practical bounds, one has to develop statistical tools. 

\subsection{Bounding the $n$ round min-entropy}
With the concepts defined above, we are now able to formulate a probabilistic statement on the min-entropy of the outputs obtained after a sequence of $n$ Bell tests. Most of this section is a reformulation, adapted to our case, of the results first presented in \cite{Pironio2010Random}, corrected in \cite{Pironio2013SecurityRandomness, Fehr2013Security}, and extended in \cite{Nieto2018Device}. Let us fix a behaviour $P_{\mathbf{A}\mathbf{B}|\mathbf{X}\mathbf{Y}}$, an i.i.d. input distribution $\pi_{xy}$, and a Bell expression $\mathcal{I}$. Then the formal statement reads:

\begin{thm} Let $\{J_m | m \in [0,M] \}$ be a sequence of $M+1$ Bell violation thresholds, with $I_{\mathcal{L}}=J_0 < J_1 < ... < J_M=I_{\mathcal{Q}}^+$. Let $\lambda_m$ be the event that the estimated Bell violation $\hat{I}$ falls between the thresholds $J_m$ and $J_{m+1}$, and let $\mathds{P}_{\tilde{P}}(\lambda_m)$ be the probability that this event occurs according to some distribution $\tilde{P}_{\mathbf{A}\mathbf{B}\mathbf{X}\mathbf{Y}}$. Let $\epsilon$ and $\epsilon'$ be two positive parameters. Then the true distribution $P_{\mathbf{A}\mathbf{B}\mathbf{X}\mathbf{Y}}$ is $\epsilon$-close to a distribution $\tilde{P}_{\mathbf{A}\mathbf{B}\mathbf{X}\mathbf{Y}}$ such that exactly one of these two statements holds:
\begin{enumerate}
\item $\mathds{P}_{\tilde{P}}(\lambda_m) \leq \epsilon'$ \label{Statement1},
\item $\Hmin(\mathbf{A},\mathbf{B}|\mathbf{X},\mathbf{Y}, \lambda_m)_{\tilde{P}_{\mathbf{A}\mathbf{B}\mathbf{X}\mathbf{Y}}} \geq \\ 
\hspace*{\fill} n H_{\mathcal{I}}^{\chi} (J_m - \mu) -\gamma(\mathbf{x})\eta -\log_2 \frac{1}{\epsilon'}$ \label{Statement2},
\end{enumerate}
where 
\begin{align}
\mu & =\nu\sqrt{\frac{2}{n}\ln{\frac{1}{\epsilon}}}, \\
\nu & =\max\{\max_{a,b,x,y} \frac{c_{abxy}}{\pi_{xy}} - I_{\mathcal{Q}}^- ,  I_{\mathcal{Q}}^+ - \min_{a,b,x,y} \frac{c_{abxy}}{\pi_{xy}}\}, \\
\gamma(\mathbf{x}) & = n - \sum_{j=1}^{n}{\mathds{1}_{\chi}(x_j)}, \\ 
\eta & = \max \{ H^\chi_{\mathcal{I}}(I_{\mathcal{Q}}^+), H^\chi_{\mathcal{I}}(I_{\mathcal{Q}}^-) \},
\end{align}
and $\mathds{1}_{\chi}(x_j)$ is the indicator function, which returns 1 if $x_j\in{\chi}$ and vanishes otherwise.
\label{Theorem}
\end{thm}

The above theorem is equivalent to Theorem 1 of~\cite{Nieto2018Device} in the case where one considers a single Bell expression. Its proof thus follows essentially the same steps as that for Theorem 1 of~\cite{Nieto2018Device} and we refer the readers to~\cite{Nieto2018Device} for details. Taking into account only one Bell expression, however, leads to numerous simplifications in its formulation, due in particular to the monotonicity of $H_{\mathcal{I}}^{\chi} $ over $[I_{\mathcal{L}},I_{\mathcal{Q}}^+]$. In this sense, it is closer to the way it is stated in \cite{Pironio2013SecurityRandomness}. However, from \cite{Nieto2018Device}, we keep a few improvements on the parameters, and the possibility to select only a subset of inputs via $\chi$. This enables improvement on the bound in some cases where the inputs have very different output probabilities: if the RB function is significantly better for a subset of inputs $\chi$, this formulation allows to use the RB function for $\chi$ only, and corrects the bound via the penalty term $\gamma(\mathbf{x})\eta$. In that case, we have an interest in biasing the input distribution towards $\chi$, in order to reduce the effect of the term $\gamma(\mathbf{x})\eta$ and thus produce as much randomness as possible. However, the trade-off between the quality of the RB function and the number of inputs from which randomness is generated depends on the total number of runs of a given protocol. 

The bound given in the second statement of the theorem is the figure of merit that we aim at optimising in this work. Indeed, this expression depends on the choice of the Bell expression $\mathcal{I}$, and we now present a systematic approach to finding a well suited $\mathcal{I}$. 

\section{Results}  
\label{sec:Results}

We first present our new method for lower-bounding the min-entropy of the outputs of an uncharacterised Bell experiment. We then study, on a few behaviours, how the regularisation method, the size of sacrificed data, and the input distributions impact the quality of the min-entropy bound.  We conclude by giving numerical results that illustrate the efficiency of our method. 

\subsection{Optimising the Bell expression via regularisation} 

As previously mentioned, solving the dual problem of \eqref{GFull} provides the Bell expression that is optimal for certifying the randomness of the given behaviour. When given an uncharacterised pair of devices, one could thus first generate some input-output data in order to estimate the corresponding underlying behaviour. This estimate $\hat{P}$ can then be used to obtain a Bell inequality that is presumably better for witnessing the randomness generated from these devices, by computing the dual solution to the guessing probability problem. Unfortunately, as mentioned above, the guessing probability problem is only properly defined over the set of quantum behaviour $\mathcal{Q}$, or one of its NPA relaxation sets $\mathcal{Q}_k$, or over the set of no-signalling behaviours. On the other hand, there is no guarantee that the behaviour built from the observed frequencies $\hat{P}$ belong to any of these sets: $\hat{P}$ is on the contrary almost always signalling, even if the underlying behaviour is not, due to finite statistics. In this case, problem \eqref{GFull} will be infeasible. 

We now introduce our method to circumvent this problem, using the tools developed in \cite{Lin2018Device}. The authors provide a set of tools to regularise the estimated behaviour $\hat{P}$ to one of the NPA sets $\mathcal{Q}_k$. It consists in minimising a norm-based metric or a statistical distance between $\hat{P}$ and $\mathcal{Q}_k$, the desired relaxation set, and taking the unique minimiser as the regularised behaviour $\Preg$. In this work, we employ two methods considered therein. The first one corresponds to minimising a statistical distance, namely the conditional Kullback-Leibler (KL) divergence \cite{Kullback1951Information, Cover2006Elements}, and is defined in the following way: 

\begin{equation}
P_{\text{ML}}(\hat{P}) = \argmin_{P \in \mathcal{Q}_k} D_{\text{KL}}(\hat{P}||P),
\label{RegML}
\end{equation}
where 
\begin{equation*}
D_{\text{KL}}(\hat{P}||P)=\sum_{a,b,x,y}\frac{N_{xy}}{n}\hat{P}(a,b|x,y)\log_2\Big(\frac{\hat{P}(a,b|x,y)}{P(a,b|x,y)}\Big).
\end{equation*}
and where ML stands for `maximal likelihood'.

The second one corresponds to minimising the two-norm distance:
\begin{equation}
P_{\text{LS}}(\hat{P}) = \argmin_{P \in \mathcal{Q}_k} \sqrt{\sum_{a,b,x,y} \Big(\hat{P}(a,b|x,y)-P(a,b|x,y) \Big)^2} ,
\label{RegLS}
\end{equation}
where `LS' stands for `least-squares'. It is important to note that both these minimisations can be efficiently solved (see \cite{Lin2018Device} for details), thus making this approach operationally relevant. A detailed study of these regularisation methods, and, in particular, of their convergence to the underlying distribution, is beyond the scope of this article; we refer the readers to \cite{Lin2018Device} for information on that subject.

We can now define the following regularisation-based protocol for generating randomness from uncharacterised devices:
\begin{enumerate}[label=(\roman*)]
\item Input a number $\Nest$ of $(x,y)$ drawn from an i.i.d. uniform distribution (they can be public) and obtain the corresponding $(a,b)$ in order to estimate the behaviour 
\item From this set of data, construct the observed frequencies $\hat{P}$ and compute $\Preg$, the regularisation of $\hat{P}$ (where $\Preg$ can be either $P_{\text{ML}}(\hat{P})$ or $P_{\text{LS}}(\hat{P})$)
\item Solve the corresponding optimisation problem $G^{\chi}_{full}(\Preg)$ for different $\chi$ and select $\chi$ accordingly (see below for further details)
\label{ProtocolSolveG}
\item Extract the optimal Bell expression $\mathcal{I}$ from the dual \label{ProtocolBestBI}
\item Input a number $\Nraw$ of $(x,y)$, drawn according to a distribution $\PXYchi$ (they can be public), obtain the corresponding $(a,b)$, and compute the observed Bell violation $\hat{I}$ \label{ProtocolRawKey}
\item Apply Theorem \ref{Theorem} to lower-bound the min-entropy of the raw set of data $(a_i,b_i,x_i,y_i)_{i\in\{1,\Nraw\}}$ \label{ProtocolMinEntropy}
\end{enumerate}

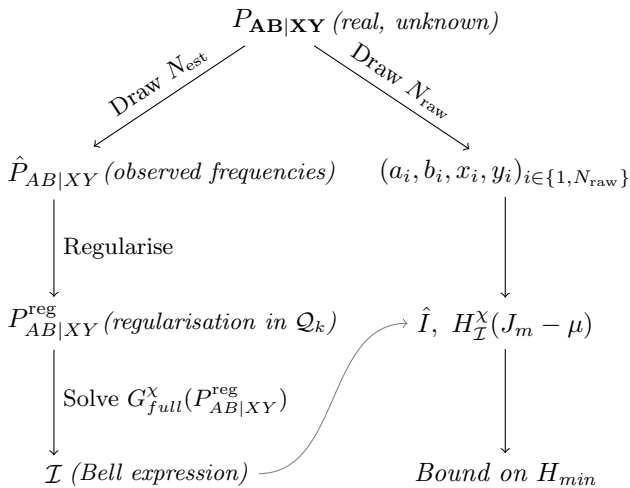
\begin{figure}
\begin{tikzpicture}
\node (a) at (0,0) {$P_{\mathbf{A}\mathbf{B}|\mathbf{X}\mathbf{Y}}$} ;
\node (b) at (1.8,0)  {\small{\textit{(real, unknown)}}};
\node (c) at (-3,-2) {$\hat{P}_{AB|XY}$ } ; 
\node (d) at (-0.8,-2) {\small{\textit{(observed frequencies)}}} ; 
\node (e) at (-3,-4) {$\Preg$} ; \node (f) at (-0.8,-4) {\small{\textit{(regularisation in $\mathcal{Q}_{k}$)}}} ; 
\node (g) at (-3,-6) {$\mathcal{I}$} ; 
\node (h) at (-1.6,-6) {\small{\textit{(Bell expression)}}}  ; 
\draw[->] (a)--(c) node[midway, above, sloped] {\small{Draw $\Nest$}};
\draw[->] (c)--(e) node[midway,right] {\small{Regularise}};
\draw[->] (e)--(g) node[midway,right] {\small{Solve $G^{\chi}_{full}(\Preg)$}};
\node (i) at (3,-2) {$(a_i,b_i,x_i,y_i)_{i\in\{1,\Nraw\}}$};
\node (j) at (3,-4) {$\hat{I}, \ H_{\mathcal{I}}^{\chi}(J_m - \mu) $};
\node (k) at (3,-6) {\textit{Bound on $\Hmin$}};
\draw[->] (a) -- (i) node[midway, above, sloped] {\small{Draw $\Nraw$}};
\draw[->, gray] (h.east) to [out=0, in=180] (j.west);
\draw[->] (i)--(j); 
\draw[->] (j) -- (k); 
\end{tikzpicture}
\caption{Schematic representation of our protocol: the user draws $\Nest$ bits from the unknown underlying behaviour, collects the frequencies and regularises them to obtain an estimate that lies in one of the NPA sets. The dual of the corresponding guessing probability problem provides a Bell inequality that is then used to quantify the min-entropy of the sequence of $\Nraw$ bits.} 
\label{Protocol}
\end{figure}

We now make a few observations on this protocol, which is summarised in Figure \ref{Protocol}. 
$\chi$ is chosen at step \ref{ProtocolSolveG}, thanks to $\Preg$. Indeed, $\Preg$ reveals some information about the underlying behaviour. One might thus intuitively do the following: compute the values of $G^{(x,y)}_{full}(\Preg)$ for all the inputs, and decide accordingly; if the value is roughly the same for all $(x,y)$, one would choose $\chi=\{0,1\}^2$;  if one input pair $(x^*,y^*)$ yields a much lower guessing probability, one would choose $\chi=(x^*,y^*)$. However, if $\Nraw$ is not big enough, $\chi=\{0,1\}^2$ is likely to result in a better min-entropy bound in any case, as our results show. 

The optimised Bell expression $\I$ obtained in step \ref{ProtocolBestBI} may not be unique and the different possible representations of $\I$ are only artefacts of numerical computations. However, the choice of a representative for $\I$ matters, since two physically equivalent representations can lead to different statistical estimates~\cite{Renou2017Inequivalence}, and thus to distinct lower bounds on the min-entropy. In order to avoid such effects, we use the unique representation introduced in~\cite{Renou2017Inequivalence}, by setting the signalling part to zero (see~\cite{Renou2017Inequivalence} for details). Note that, after this step, the regularised distribution no longer plays a role, and the min-entropy bound associated to that Bell inequality is valid independently of which regularisation method was used.

In step \ref{ProtocolRawKey}, we assume that the specific distributions $\PXYchi$ can be generated using some freely available resource. If this is the case, one might consider that the task of randomness generation is already achievable, and we might then call our primitive `randomness expansion', rather than `randomness generation'. However, the input randomness can be public: it needs to be random to anyone beforehand, but it can be accessed by anyone after it is produced. Conversely, the output randomness is private: its value resides in the fact that it is only accessible to the user. We can thus refer to this process as `private random bits generation'.

In step \ref{ProtocolMinEntropy}, in order to apply Theorem \ref{Theorem}, we need to know the quantum bounds $I_{\mathcal{Q}}^+$ and $I_{\mathcal{Q}}^-$. We approximate these bounds with the extrema over an NPA set, so that they can be easily computed. Moreover, we only bound the min-entropy of the data generated in step \ref{ProtocolRawKey}. Indeed, it is essential that the set of data used for the estimation be different from the one for which the bound on the min-entropy is derived: the statistical analysis of the data cannot depend on the data itself. This implies that, contrarily to \cite{Nieto2018Device}, our method requires that part of the data is used only for parameter estimation, and then thrown away. 

Finally, note that even though the regularisation method described in \cite{Lin2018Device} is meaningful only when the underlying distribution $P_{\mathbf{A}\mathbf{B}|\mathbf{X}\mathbf{Y}}$ is i.i.d., the derivation of the bound on the min-entropy does not rely on this assumption. For this reason, the probabilistic statement that we obtain via our method will still be valid, even if $P_{\mathbf{A}\mathbf{B}|\mathbf{X}\mathbf{Y}}$ is not i.i.d.. In this case, the Bell expression that we obtain might be inadequate, which might result in a trivial lower bound on the min-entropy (that equals to zero), but it will not result in an overestimation of the min-entropy of the raw data. In this sense, the optimisation method might become irrelevant, but the security analysis will not be compromised.

\subsection{Tuning the parameters}
\label{sec:Tuning}

In order to adjust the parameters of our protocol, we simulate some pairs of devices, by generating for each one a random state $\rho$ and some random measurements $\{M^\mathrm{A}_{a|x}\}_a$ and $\{M^\mathrm{B}_{b|y}\}_b$. The random states are picked at random in the space of two qubit pure states via their Schmidt decomposition, and the random measurements are generated via their associated rank-1 projectors, picked at random on the Bloch sphere. 

We then compute the associated behaviour:
\begin{equation}
P_{AB|XY}(ab|xy)= \text{Tr}[\rho\, M^A_{a|x}\otimes M^B_{b|y}].
\end{equation}

To ensure that the obtained behaviours are non-local, we compute their associated values $\mathcal{I}_{\text{CHSH}}(P_{AB|XY})$ of the Clauser-Horne-Shimony-Holt (CHSH) inequality \cite{Clauser1969Proposed}: 
\begin{equation}
\mathcal{I}_{\text{CHSH}}(P_{AB|XY})=\sum_{x,y,a,b} (-1)^{xy+a+b}P_{AB|XY}(ab|xy),
\end{equation}
and discard those for which $\mathcal{I}_{\text{CHSH}}(P_{AB|XY}) \leq 2$. We then construct the corresponding $\Ntot$-round behaviour using $P_{AB|XY}$ in an i.i.d. way, i.e., 
\begin{equation}
P_{\mathbf{A}\mathbf{B}|\mathbf{X}\mathbf{Y}}(\mathbf{a}\mathbf{b}|\mathbf{x}\mathbf{y})=\prod_{i=1}^{\Ntot} P_{AB|XY}(a_ib_i|x_iy_i).
\end{equation}

We set $\Ntot=\Nest+\Nraw = 10^8$, in accordance with the state-of-the-art experimental demonstration of device-independent randomness generation~\cite{Bierhorst2018Experimentally}. We then conduct a detailed study of four of these random behaviours, to heuristically fix three crucial parameters of our protocol: 
\begin{itemize}
\item the regularisation method,
\item the number of rounds used for the estimation $\Nest$,
\item the inputs subset used to generate randomness $\chi$,
\end{itemize}

Based on the data we obtained, presented in Appendix~\ref{app:Tuning}, we decided to set:
\begin{itemize}
\item $P_{AB|XY}^{\mathrm{reg}}= P_{ML}$,
\item $\Nest=10^6$,
\item  $\chi=\{0,1\}^2$
\end{itemize}

The graphs that corroborate these decisions can be found in Appendix~\ref{app:Tuning}. Before we give the results of several simulations that illustrate the efficiency of our protocol, note that, when one sets $\Ntot=10^8$, generating randomness from only one input pair (i.e., setting $\chi=(x^*,y^*)$) does not usually result in higher min-entropy bounds than when one sets $\chi=\{0,1\}^2$. The same effect can be observed in the simulations carried out by the authors in~\cite{Nieto2018Device}. It is not surprising: in order to obtain a good min-entropy rate when certifying randomness from only one input pair, one should bias the input distribution towards that pair as much as possible. However, in order to obtain a reliable estimate of the Bell violation, one should evaluate it with many occurrences of each possible input. These two assertions are in an apparent contradiction, and they can both hold simultaneously only if $\Ntot$ is high enough. It seems that, for most behaviours, $\Ntot=10^8$ is not sufficient. We however checked that, when $\Ntot$ is sufficiently big, our method provides better min-entropy bounds for $\chi=(x^*,y^*)$ than for $\chi=\{0,1\}^2$. The corresponding graph can be found in Appendix~\ref{app:ChiOne}. 

\subsection{Numerical results}
\label{NumRes}

Our figure of merit is the comparison between the min-entropy bound obtained from our protocol, denoted $\Hmin$ in the following, and the one obtained from a direct evaluation of the CHSH inequality, $\Hmin^{\rm{CHSH}}$. We generate 50 behaviours at random (in the same way as described above) and run 500 simulations for each of them. To compute the lower bound on $\Hmin$, one should set $n=\Nraw$ in Theorem \ref{Theorem}, whereas for $\Hmin^{\rm{CHSH}}$, $n=\Ntot > \Nraw$, as no estimation is required.\footnote{It might seem necessary to also first sacrifice a part of the data to determine which among the 8 representatives of the CHSH inequality is violated. This is however unnecessary as any given behaviour can violate at most one representative of the CHSH inequality (see page 2 of the Supplementary Material to~\cite{Liang:PRL:2010}), which can be determined by evaluating the min-entropy bound of all different representatives of the CHSH inequality.}

The parameters of the bound of Theorem \ref{Theorem} are set as follows: we fix $\epsilon=\epsilon'=10^{-6}$, we divide the interval $[I_{\mathcal{L}} , I_{\mathcal{Q}}^+]$ in $M+1=1000$ segments of the same length, and we use the level 2 of the hierarchy defined in~\cite{Moroder2013Device} (i.e., local level 2 defined in~\cite{Vallins2017}) for the regularisation and the guessing probability problems. We then compute the corresponding min-entropy rate by dividing these values by $\Ntot$ in both cases. We also computed $-\log_2(G^{\chi}_{full}(P_{AB|XY}))$, which corresponds to the maximal achievable min-entropy rate. To show that it is worth sacrificing part of the data for estimation, we then compared these three quantities. The results are presented in Fig. \ref{fig:10e8}.

\newsavebox{\mybox}
\savebox{\mybox}{
    \begin{tikzpicture}
      \begin{axis}[
        xmin=0,
        xmax=50.5,
        xtick=\empty,
        ymin=0,
        ymax=2.5,
        ytick={0,0.5,1,1.5,2,2.5},
	grid = major,
	axis background/.style={fill=white} , set layers, cell picture=true
        ]
        \addplot [color=black, draw=none, thick, mark=asterisk]
        table[row sep=crcr]{
1	0.990848687291579\\
2	0.989840250679943\\
3	0.989868713404485\\
4	0.989672607343848\\
5	0.991828134714621\\
6	0.989687631845657\\
7	0.99775764186236\\
8	0.99369951076546\\
9	1.00534932003412\\
10	1.00931159716925\\
11	1.0319347775844\\
12	1.02791516870672\\
13	1.02936675708063\\
14	1.02527066043776\\
15	1.04967826513898\\
16	1.03967890409824\\
17	1.06254852909612\\
18	1.03226567312041\\
19	1.05030364274032\\
20	1.05811715817716\\
21	1.06121655294768\\
22	1.05264604385807\\
23	1.0687417501135\\
24	1.08217936422888\\
25	1.08988128232927\\
26	1.08508836755344\\
27	1.03209706386115\\
28	1.11713051411684\\
29	1.10236039282319\\
30	1.14538536422961\\
31	1.10629143855303\\
32	1.14053075597109\\
33	1.16911575041434\\
34	1.18166367922354\\
35	1.20273824177717\\
36	1.28195836436737\\
37	1.25715123005844\\
38	1.29342591267224\\
39	1.18384182510108\\
40	1.24461496883788\\
41	1.28948094044959\\
42	1.39587117689041\\
43	1.32552989273358\\
44	1.43182835015018\\
45	1.41570546591348\\
46	1.58261119954058\\
47	1.67172998409216\\
48	1.6306764773446\\
49	1.31714913141478\\
50	0.00279102803467605\\
        };
        \addplot[name path=A, draw=none, color=red, mark=o]
        table[row sep=crcr]{
1	1.02301351215486\\
2	1.02396504069727\\
3	1.02495166337562\\
4	1.02637542156434\\
5	1.02717650406852\\
6	1.03756212959124\\
7	1.04289272579622\\
8	1.04831474747437\\
9	1.05728812433414\\
10	1.06122640329803\\
11	1.06267358464216\\
12	1.06894066712846\\
13	1.06940959556623\\
14	1.07178536016224\\
15	1.0739615420659\\
16	1.08154010793909\\
17	1.08675708170576\\
18	1.08855019979664\\
19	1.09033920071769\\
20	1.09162363541352\\
21	1.09237097025272\\
22	1.09460879526421\\
23	1.10018654291637\\
24	1.11946675195672\\
25	1.13421297600194\\
26	1.13505094197572\\
27	1.13926595808251\\
28	1.14416803603698\\
29	1.15425698912241\\
30	1.1866917176338\\
31	1.20826374375302\\
32	1.21012812172801\\
33	1.23186838518755\\
34	1.24126970760298\\
35	1.26043416989659\\
36	1.32685298948722\\
37	1.33299469320543\\
38	1.33341413236496\\
39	1.34050917695519\\
40	1.36224882216975\\
41	1.40653031129862\\
42	1.43703249246313\\
43	1.45961478647705\\
44	1.48653698065125\\
45	1.71112918136554\\
46	1.72891177035038\\
47	1.79672385876672\\
48	1.80676691022249\\
49	1.97841405945132\\
50	2.38689948288473\\
        };
        \addplot [name path=B, black, domain={1:50.5}] {1};
        \addplot[gray, fill opacity=0.6]fill between[of=A and B, soft clip={domain=1:50}];
      \end{axis}
    \end{tikzpicture}}

\begin{figure}
    \begin{tikzpicture}
      \begin{axis}[
        xmin=0,
        xmax=49.5,
        xtick=\empty,
        ymin=0.9,
        ymax=2,
        ylabel=$\langle \Hmin / \Hmin^{\rm{CHSH}} \rangle$, 
        ytick={1,1.25,1.5,1.75, 2},
        legend style={at={(axis cs:45,1.968)}}, 
        grid=major
        ]
        \addlegendimage{only marks, mark=asterisk, black, thick}
        \addlegendimage{only marks, mark=o, red}
        \addplot [color=black, draw=none, thick, mark=asterisk]
        table[row sep=crcr]{
1	0.990848687291579\\
2	0.989840250679943\\
3	0.989868713404485\\
4	0.989672607343848\\
5	0.991828134714621\\
6	0.989687631845657\\
7	0.99775764186236\\
8	0.99369951076546\\
9	1.00534932003412\\
10	1.00931159716925\\
11	1.0319347775844\\
12	1.02791516870672\\
13	1.02936675708063\\
14	1.02527066043776\\
15	1.04967826513898\\
16	1.03967890409824\\
17	1.06254852909612\\
18	1.03226567312041\\
19	1.05030364274032\\
20	1.05811715817716\\
21	1.06121655294768\\
22	1.05264604385807\\
23	1.0687417501135\\
24	1.08217936422888\\
25	1.08988128232927\\
26	1.08508836755344\\
27	1.03209706386115\\
28	1.11713051411684\\
29	1.10236039282319\\
30	1.14538536422961\\
31	1.10629143855303\\
32	1.14053075597109\\
33	1.16911575041434\\
34	1.18166367922354\\
35	1.20273824177717\\
36	1.28195836436737\\
37	1.25715123005844\\
38	1.29342591267224\\
39	1.18384182510108\\
40	1.24461496883788\\
41	1.28948094044959\\
42	1.39587117689041\\
43	1.32552989273358\\
44	1.43182835015018\\
45	1.41570546591348\\
46	1.58261119954058\\
47	1.67172998409216\\
48	1.6306764773446\\
49	1.31714913141478\\
        };
        \addlegendentry{Protocol};
        \addplot[name path=A, draw=none, color=red, mark=o]
        table[row sep=crcr]{
1	1.02301351215486\\
2	1.02396504069727\\
3	1.02495166337562\\
4	1.02637542156434\\
5	1.02717650406852\\
6	1.03756212959124\\
7	1.04289272579622\\
8	1.04831474747437\\
9	1.05728812433414\\
10	1.06122640329803\\
11	1.06267358464216\\
12	1.06894066712846\\
13	1.06940959556623\\
14	1.07178536016224\\
15	1.0739615420659\\
16	1.08154010793909\\
17	1.08675708170576\\
18	1.08855019979664\\
19	1.09033920071769\\
20	1.09162363541352\\
21	1.09237097025272\\
22	1.09460879526421\\
23	1.10018654291637\\
24	1.11946675195672\\
25	1.13421297600194\\
26	1.13505094197572\\
27	1.13926595808251\\
28	1.14416803603698\\
29	1.15425698912241\\
30	1.1866917176338\\
31	1.20826374375302\\
32	1.21012812172801\\
33	1.23186838518755\\
34	1.24126970760298\\
35	1.26043416989659\\
36	1.32685298948722\\
37	1.33299469320543\\
38	1.33341413236496\\
39	1.34050917695519\\
40	1.36224882216975\\
41	1.40653031129862\\
42	1.43703249246313\\
43	1.45961478647705\\
44	1.48653698065125\\
45	1.71112918136554\\
46	1.72891177035038\\
47	1.79672385876672\\
48	1.80676691022249\\
49	1.97841405945132\\
        };
        \addlegendentry{Optimal};        
        \addplot [name path=B, black, domain={1:49}] {1};
        \addplot[gray, fill opacity=0.6]fill between[of=A and B, soft clip={domain=1:50}];
        \draw (axis cs: 15,1.69)node[scale=0.5]{\usebox{\mybox}};
      \end{axis}
    \end{tikzpicture}
\caption{Black asterisk: ratio between the rate obtained via our protocol and via the direct use of the CHSH inequality. Red circle: ratio between the maximal achievable min-entropy and the rate obtained via the direct use of the CHSH inequality. The inset contains all 50 simulations, including the single instance from which no randomness is certified (see explanation in the main text). This exceptional point is removed from the main plot so that the remaining (successful) cases can be examined more closely.}
\label{fig:10e8}
\end{figure}
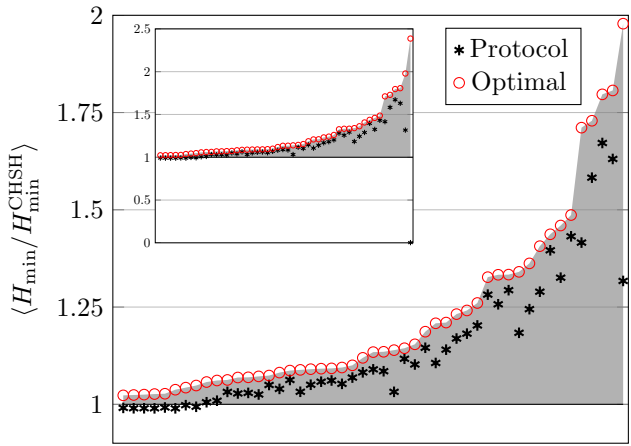

In this figure, we plot the ratios between the min-entropy rates for $\Hmin$ and $\Hmin^{\rm{CHSH}}$ for every simulated pairs of devices, as well as the ratios between the maximal achievable rate $-\log_2(G^{\chi}_{full}(P_{AB|XY}))$ and $\Hmin^{\rm{CHSH}}$. For clarity, we sorted them in ascending order of the latter. We highlighted in grey the areas between the line $y=1$, where the amount of randomness given by our protocol is the same as using CHSH inequality, and the curves connecting the optimal ratios. Our protocol is good whenever a point falls in this area. Indeed, it means that, despite the $\Nest$ bits that were thrown away, we obtain a higher bound on the min-entropy than if we had simply used the CHSH inequality on all the bits. 

We observe that our method performs well in $98\%$ of the simulations, in the following sense: when the optimal rate is nearly achieved with the CHSH inequality (i.e., the CHSH inequality gives a bound that is above $95\%$ of the optimal rate), so does our method; when the CHSH inequality does not achieve the optimal rate, our method performs significantly better (with rates up to 1.6 times more) in all but one case.  

Before we move on to concluding remarks, let us explain what happened with the last point of our simulations, from which no randomness can be certified via our protocol. The corresponding underlying behaviour has a low CHSH value, and the optimal Bell inequality is such that the gap between the local bound and the quantum bound is very small. This seems to indicate that this behaviour is of the kind presented in~\cite{Acin2012Randomness}, i.e., it is almost local, but also close to the border of the quantum set. The authors of~\cite{Acin2012Randomness} proved that, in theory, a lot of randomness could be certified from such behaviours, as can be observed by the corresponding red circle in Fig.~\ref{fig:10e8}. However, those behaviours are not good from a practical point of view: the small gap between the local and the quantum bounds of their associated optimal Bell inequality requires that the confidence interval on the estimated Bell violation $J_m - \mu$ be very small. If not, i.e., if $J_m - \mu$ is smaller than the local bound, no Bell violation can be observed, and thus no randomness can be certified. This is the case for that point of our simulations. 

\section{Conclusion and future works}
\label{sec:Conclusion}

We presented a simple method to optimise the lower bound derived in~\cite{Nieto2018Device} on the min-entropy produced by a sequence of Bell tests. It improves the analysis of the data collected from these Bell tests, without requiring changes in the current implementations of device-independent randomness generation. It consists in estimating the underlying behaviour of the black boxes, via the regularisation method given in \cite{Lin2018Device}. We then tuned the parameters of this protocol via a heuristic method. We concluded that, when one regularises some data for randomness generation, one should always use the maximal likelihood method (the authors observed the same effect for another figure of merit, the negativity~\cite{Moroder2013Device}, in~\cite{Lin2018Device}), one can sacrifice up to $1\%$ of the data for estimation, and that, for the device-independent randomness generation experiments that can be performed at the moment (i.e., with $\Ntot=10^8$), one should generally use the worst case RB function (i.e., the one that bounds the randomness for all inputs). We then carried out numerical simulations that illustrate the efficiency of this method. Comparing it with other existing methods on real experimental data, such as the one presented in~\cite{Liu2018Device}, would also be insightful. However, the publicly available data for these experiments is not sufficient at the moment, as it only reports the total number of occurrences of all inputs and outputs, while we would need to split the sequentially obtained data into two sets, one for estimation and one for randomness generation.

We now describe two possible lines of investigation that follow from this work. The first one would be to take into account more factors in the optimisation of the lower bounds on the min-entropy. For instance, one could generate randomness from two or three subsets of inputs pairs, instead of considering only one or all of them as we did here. One could also tune $\PXYchi$ in a more precise way, as a function of the total number of rounds $\Ntot$ and of the differences between the guessing probabilities for each input pair.  Finally, the RB function is a key element in the derivation of the bound. We used here the one introduced in \cite{Nieto2018Device}. However, there are other ways to compute a function that satisfies both requirements \ref{Req1} and \ref{Req2} needed for an RB function, such as the one introduced in \cite{Bancal2014More}. Being able to compute the RB function that is tight would entail an improvement on the min-entropy bound. 

The second one is related to the power given to the adversary. Our results hold in a trusted provider scenario, where our protocol allows for correcting noise and deterioration in the apparatuses, and in an adversarial scenario where the adversary holds only classical-side information. Adapting it to the case of an adversary with quantum side information would provide a min-entropy bound valid in the most general scenario. This could be achieved using a recent result, the entropy accumulation theorem~\cite{Dupuis2016Entropy}. Based on that result, a bound was derived on the $n$-round smooth min-entropy against an adversary with quantum side information~\cite{Arnon2018Practical}. However, this bound is based on the CHSH inequality. Deriving such a bound for other inequalities might be a hard task. We took a different approach here, that consists in optimising the amount of randomness that is generated by tailoring the Bell inequality to a specific case. This, in turn, led us to consider only classical side information. If one could adapt the results of~\cite{Dupuis2016Entropy, Arnon2018Practical} to any Bell inequality, one would be able to guarantee the security of our protocol in the most general scenario. 

\section*{Acknowledgements}

We thank Jean-Daniel Bancal and Yanbao Zhang for useful discussions. Support from the ERC CoG QITBOX, the AXA Chair in Quantum Information Science, the Spanish MINECO (QIBEQI FIS2016-80773-P and Severo Ochoa SEV-2015-0522), the Generalitat de Catalunya (CERCA Program SGR 1381), the Fundaci\'o Privada Cellex, the Ministry of Science and Technology, Taiwan (Grants No. 104-2112-M-006-021-MY3 and No. 107-2112-M-006-005-MY2), and the Swiss National Science Foundations (SNSF) [Grant number P2GEP2$\_$162060 (Early Postdoc Mobility Grant)] is acknowledged. BB acknowledges support from the Secretaria d'Universitats i Recerca del Departament d'Economia i Coneixement de la Generalitat de Catalunya and the European Social Fund - FEDER.  

\bibliographystyle{apsrev4-1}
\bibliography{RandomnessRegularisation.bib}

\begin{widetext}
\appendix
\section{Tuning the parameters}
\label{app:Tuning}

We present here the analysis that we conducted in order to tune the parameters of our protocol. We first generated four random distributions, in the same way as explained in the main text, and computed the min-entropy rates for varying $\Nest$, to see how many bits should be sacrificed for estimation. We set the parameters of the bound in the same way as presented the main text, i.e., we fix $\epsilon=\epsilon'=10^{-6}$, we divide the interval $[I_{\mathcal{L}} , I_{\mathcal{Q}}^+]$ in $M+1=1000$ segments of the same length, we use the NPA local level 2 \cite{Moroder2013Device} for the regularisation and the guessing probability problems, we set $\Ntot=10^8$, and we run 500 simulations for each point. The inputs distribution for the estimation phase is always uniform. We compute the average min-entropy rates $\langle \Hmin / \Ntot \rangle$ as a function of $\log_{10} \Nest$ for both regularisation methods ML and LS, and with two possible choices for $\chi$: $\chi_{\textrm{all}}=\{0,1\}^2$ and $\chi_{\textrm{one}}=(x^*,y^*)$, where $(x^*,y^*)$ is the most random input pair, i.e. the one that yields the highest RB function. In that case, we set the input distribution to $P_{XY}(x^*,y^*)=\pi_{x^*y^*}=0.9$ (and uniform on the other inputs). The results are presented in Figure \ref{fig:nest}. 

\begin{figure}
\begin{subfigure}{0.49\linewidth}
\begin{tikzpicture}
\begin{axis}[
/pgfplots/tick scale binop=\times,
/pgf/number format/.cd, fixed,
xmin=2,
xmax=7,
ymin=0,
ymax=0.2,
height=7cm,
xlabel=$\log_{10}\Nest$, 
ylabel=$\langle \Hmin / \Ntot \rangle$, 
legend columns=2,
legend style= {at = {(0.75,0.52)}, anchor=south east}]
\addplot[green!60!black, ultra thick, domain=2:7, samples=10] {0.179261783183499};
\addplot[purple, domain=2:7, samples=10, double] {0.0971840172020431};
\addplot [color=green!60!black, mark=triangle*]
  table[row sep=crcr]{
2	0\\
3	1.46911311377957e-07\\
4	0\\
5	0\\
6	0\\
7	0\\
};
\addplot [color=purple, mark=triangle, dashed, every mark/.append style={solid}]
table[row sep=crcr]{
2	0.000310407353349736\\
3	0.0241590484456001\\
4	0.0884301865706189\\
5	0.0918375637039692\\
6	0.0913019250224349\\
7	0.0828145395896474\\
};
\addplot [color=green!60!black, mark=*]
table[row sep=crcr]{
2	0\\
3	0\\
4	0\\
5	0\\
6	0\\
7	0\\
};
\addplot[color=purple, mark=o, dashed, every mark/.append style ={solid}]
table[row sep=crcr]{
2	0.00142337023647928\\
3	0.0153851604814174\\
4	0.0533138697423814\\
5	0.0898614574399359\\
6	0.0910956418156531\\
7	0.082806451512816\\
};
\addplot [color=blue, densely dotted]
table[row sep=crcr]{
2	0.0710745526255352\\
3	0.0710488456137799\\
4	0.0710390967053013\\
5	0.0710550558265796\\
6	0.0710586025529162\\
7	0.0710577193130426\\
};
\addlegendentry{Opt, $\chi_{\textrm{one}}$};
\addlegendentry{Opt, $\chi_{\textrm{all}}$};
\addlegendentry{ML, $\chi_{\textrm{one}}$};
\addlegendentry{ML, $\chi_{\textrm{all}}$};
\addlegendentry{LS, $\chi_{\textrm{one}}$};
\addlegendentry{LS, $\chi_{\textrm{all}}$};
\addlegendentry{CHSH};
\end{axis}
\end{tikzpicture}
\caption{$I_{\rm{CHSH}}=2.1579$}
\end{subfigure}
\begin{subfigure}{0.49\linewidth}
\begin{tikzpicture}
\begin{axis}[
/pgfplots/tick scale binop=\times,
/pgf/number format/.cd, fixed,
xmin=2,
xmax=7,
ymin=0,
ymax=0.25,
height=7cm,
xlabel=$\log_{10}\Nest$,
ylabel=$\langle \Hmin / \Ntot \rangle$]
\addplot[green!60!black, ultra thick, domain=2:7, samples=10] {0.244277707631205};
\addplot [color=green!60!black, mark=triangle*]
table[row sep=crcr]{
2	3.02912913058252e-05\\
3	0.0283126739517106\\
4	0.0556142718451685\\
5	0.0689971024227457\\
6	0.0682200709336463\\
7	0.0604477121424163\\
};
\addplot [color=green!60!black, mark=*]
table[row sep=crcr]{
2	0.00254089330733994\\
3	0.0226635506708703\\
4	0.0444277558846007\\
5	0.0629391003890956\\
6	0.0683238363210561\\
7	0.060349869874439\\
};
\addplot[purple, domain=1:10, samples=10, double] {0.138997073548588};
\addplot [color=purple, mark=triangle, dashed,  every mark/.append style={solid}]
table[row sep=crcr]{
2	0.000715129149517622\\
3	0.0826521349307003\\
4	0.1324792552706\\
5	0.134813698006477\\
6	0.133812767672557\\
7	0.121504499480069\\
};
\addplot[color=purple, mark=o, dashed,  every mark/.append style={solid}]
table[row sep=crcr]{
2	0.00886585648698709\\
3	0.0689163101626291\\
4	0.11786051075489\\
5	0.134144047349246\\
6	0.133761969033668\\
7	0.121500434977612\\
};
\addplot [color=blue, densely dotted]
  table[row sep=crcr]{
2	0.129981846329994\\
3	0.129973407799113\\
4	0.129999755654464\\
5	0.12997973970217\\
6	0.129986056573974\\
7	0.12998817372232\\
};
\end{axis}
\end{tikzpicture}
\caption{$I_{\rm{CHSH}}=2.2587$}
\end{subfigure}
\begin{subfigure}{0.49\linewidth}
\begin{tikzpicture}
\begin{axis}[
/pgfplots/tick scale binop=\times,
/pgf/number format/.cd, 
fixed,
xmin=2,
xmax=7,
ymin=0,
ymax=0.85,
height=7cm,
xlabel=$\log_{10}\Nest$,
ylabel=$\langle \Hmin / \Ntot \rangle$]
\addplot[green!60!black, ultra thick, domain=2:7, samples=10] {0.809928739754258};
\addplot [color=green!60!black, mark=triangle*]
table[row sep=crcr]{%
2	0.000983367848945279\\
3	0.0608774514987824\\
4	0.169681683248277\\
5	0.27278342367971\\
6	0.285657131963991\\
7	0.246489414144667\\
};
\addplot [color=green!60!black, mark=*]
table[row sep=crcr]{%
2	0.00752641849950577\\
3	0.0431024843137594\\
4	0.0934857967519559\\
5	0.205358914015428\\
6	0.269191898737714\\
7	0.245062382360587\\
};
\addplot[purple, domain=2:7, samples=10, double] {0.385530813537734};

\addplot [color=purple, mark=triangle, dashed, every mark/.append style={solid}]
table[row sep=crcr]{%
2	0.00931310352118362\\
3	0.0918359095710416\\
4	0.29746514594192\\
5	0.356735983076397\\
6	0.355991373199978\\
7	0.322681025777578\\
};
\addplot [color=purple, mark=o, dashed,  every mark/.append style={solid}]
table[row sep=crcr]{%
  2	0.0190758133388611\\
3	0.0731197642510127\\
4	0.179059194191932\\
5	0.331651284110056\\
6	0.354764318942218\\
7	0.322528373807957\\
};
\addplot [color=blue, densely dotted]
table[row sep=crcr]{%
2	0.249480719524304\\
3	0.249462244966488\\
4	0.249462256955315\\
5	0.249473615585136\\
6	0.249445192635166\\
7	0.249484981408252\\
};
\end{axis}
\end{tikzpicture}
\caption{$I_{\rm{CHSH}}=2.4206$}
\end{subfigure}
\begin{subfigure}{0.49\linewidth}
\begin{tikzpicture}
\begin{axis}[
/pgfplots/tick scale binop=\times,
/pgf/number format/.cd, fixed,
xmin=2,
xmax=7,
ymin=0,
ymax=1.2,
height=7cm,
xlabel=$\log_{10}\Nest$, 
ylabel=$\langle \Hmin / \Ntot \rangle$]
\addplot[green!60!black, ultra thick, domain=2:7, samples=10] {1.17058210790156};
\addplot [color=green!60!black, mark=triangle*]
  table[row sep=crcr]{
2	0.0222313620781811\\
3	0.286320902952853\\
4	0.532215883176318\\
5	0.720332030826367\\
6	0.720680368174893\\
7	0.644705396504813\\
};
\addplot [color=green!60!black, mark=*]
table[row sep=crcr]{
2	0.0445819044222381\\
3	0.187842740877536\\
4	0.422827595181866\\
5	0.685598470860088\\
6	0.720735377920189\\
7	0.643647408027885\\
};
\addplot[purple, domain=2:7, samples=10, double] {0.516204956513184};
\addplot [color=purple, mark=triangle, dashed, every mark/.append style={solid}]
table[row sep=crcr]{
2	0.0455993711351456\\
3	0.307871999992392\\
4	0.436116434119498\\
5	0.491883017604196\\
6	0.494907929165384\\
7	0.449682413509711\\
};
\addplot[color=purple, mark=o, dashed, every mark/.append style ={solid}]
table[row sep=crcr]{
2	0.0732051426991653\\
3	0.25733527001402\\
4	0.377139613321347\\
5	0.480090450362299\\
6	0.494050895931655\\
7	0.449658450177291\\
};
\addplot [color=blue, densely dotted]
table[row sep=crcr]{
2	0.452400483757885\\
3	0.452414192325005\\
4	0.45240047759227\\
5	0.452423308814373\\
6	0.452418765983725\\
7	0.45242329039534\\
};
\end{axis}
\end{tikzpicture}
\caption{$I_{\rm{CHSH}}=2.6090$}
\end{subfigure}
\caption{Average min-entropy rates as a function of the size of the data that is sacrificed for estimation.}
\label{fig:nest}
\end{figure}
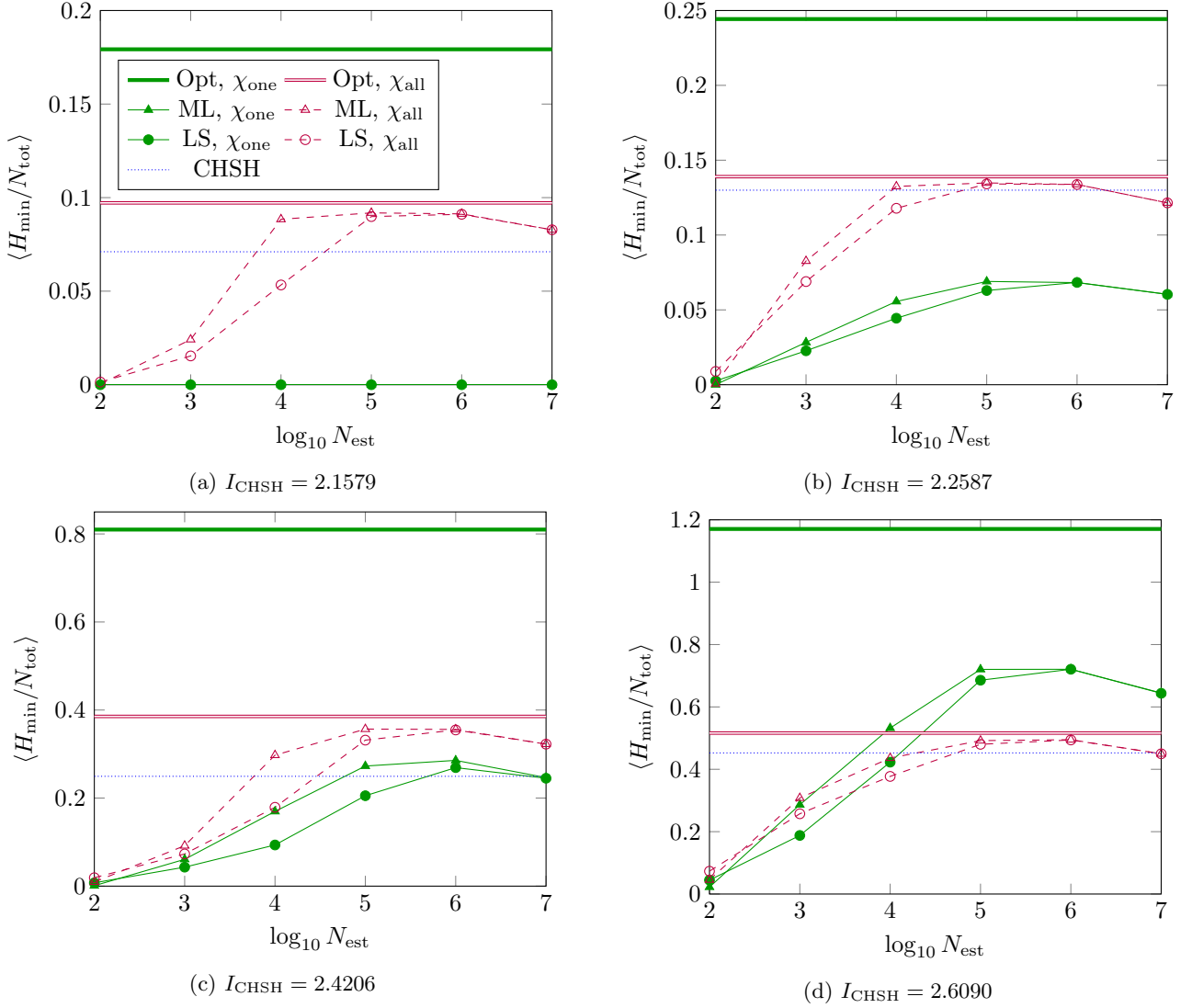

From those graphs, we deduce that setting $\Nest=10^6$, i.e., $1\%$ of the total data, is optimal. Note that, to distinguish these four distributions, we give their CHSH values $I_{\textrm{CHSH}}$. It does not mean that the CHSH inequality is the best Bell expression for certifying randomness from these behaviours: we merely give it as a way to quantify how non-local these distributions are, because it might be interesting for the reader to see that the effects we observe seem to depend on that. For instance, generating randomness from only one input seems to give an advantage only when the CHSH value is high enough. 

We then study, under the same conditions, the effect of the input in the bias distribution $\pi_{x^*y^*}$, to see if one can observe an advantage when setting $\chi=\chi_{\textrm{one}}$ instead of $\chi=\chi_{\textrm{all}}$. The results can be found in Figure \ref{fig:inputbias}.

\begin{figure}
\begin{subfigure}{0.49\linewidth}
\begin{tikzpicture}
\begin{axis}[
/pgfplots/tick scale binop=\times,
/pgf/number format/.cd, fixed,
xmin=0.25,
xmax=0.95,
ymin=0,
ymax=0.2,
height=7cm,
xlabel=$\pi_{x^*y^*}$,
ylabel=$\langle \Hmin / \Ntot \rangle$, 
legend columns =2, legend style= {at = {(0.75,0.52)}, anchor=south east}]
\addlegendentry{Opt, $\chi_{\textrm{one}}$};
\addlegendentry{Opt, $\chi_{\textrm{all}}$};
\addlegendentry{ML, $\chi_{\textrm{one}}$};
\addlegendentry{ML, $\chi_{\textrm{all}}$};
\addlegendentry{LS, $\chi_{\textrm{one}}$};
\addlegendentry{LS, $\chi_{\textrm{all}}$};
\addlegendentry{CHSH};
\addplot[green!60!black, ultra thick, domain=0.25:0.95, samples=15] {0.179261783183762};
\addplot[purple, domain=0.25:0.95, samples=15, double] {0.0971840196872175};
\addplot [color=green!60!black, mark=triangle*]
  table[row sep=crcr]{
0.25	0\\
0.3	0\\
0.35	0\\
0.4	0\\
0.45	0\\
0.5	0\\
0.55	0\\
0.6	0\\
0.65	0\\
0.7	0\\
0.75	0\\
0.8	0\\
0.85	0\\
0.9	0\\
0.95	0\\
};
\addplot [color=purple, mark=triangle, dashed,  every mark/.append style={solid}]
table[row sep=crcr]{
0.25	0.091337889915678\\
0.3	0.091337889915678\\
0.35	0.091337889915678\\
0.4	0.091337889915678\\
0.45	0.091337889915678\\
0.5	0.091337889915678\\
0.55	0.091337889915678\\
0.6	0.091337889915678\\
0.65	0.091337889915678\\
0.7	0.091337889915678\\
0.75	0.091337889915678\\
0.8	0.091337889915678\\
0.85	0.091337889915678\\
0.9	0.091337889915678\\
0.95	0.091337889915678\\
};
\addplot [color=green!60!black, mark=*]
table[row sep=crcr]{
0.25	0\\
0.3	0\\
0.35	0\\
0.4	0\\
0.45	0\\
0.5	0\\
0.55	0\\
0.6	0\\
0.65	0\\
0.7	0\\
0.75	0\\
0.8	0\\
0.85	0\\
0.9	0\\
0.95	0\\
};
\addplot[color=purple, mark=o,dashed,  every mark/.append style={solid}]
table[row sep=crcr]{
0.25	0.0911572772057908\\
0.3	0.0911572772057908\\
0.35	0.0911572772057908\\
0.4	0.0911572772057908\\
0.45	0.0911572772057908\\
0.5	0.0911572772057908\\
0.55	0.0911572772057908\\
0.6	0.0911572772057908\\
0.65	0.0911572772057908\\
0.7	0.0911572772057908\\
0.75	0.0911572772057908\\
0.8	0.0911572772057908\\
0.85	0.0911572772057908\\
0.9	0.0911572772057908\\
0.95	0.0911572772057908\\
};
\addplot [color=blue, densely dotted]
table[row sep=crcr]{
0.25	0.0710843081106756\\
0.3	0.0710843081106756\\
0.35	0.0710843081106756\\
0.4	0.0710843081106756\\
0.45	0.0710843081106756\\
0.5	0.0710843081106756\\
0.55	0.0710843081106756\\
0.6	0.0710843081106756\\
0.65	0.0710843081106756\\
0.7	0.0710843081106756\\
0.75	0.0710843081106756\\
0.8	0.0710843081106756\\
0.85	0.0710843081106756\\
0.9	0.0710843081106756\\
0.95	0.0710843081106756\\
};
\end{axis}
\end{tikzpicture}
\caption{$I_{\rm{CHSH}}=2.1579$}
\end{subfigure}
\begin{subfigure}{0.49\linewidth}
\begin{tikzpicture}
\begin{axis}[
/pgfplots/tick scale binop=\times,
/pgf/number format/.cd, fixed,
xmin=0.25,
xmax=0.95,
ymin=0,
ymax=0.25,
height=7cm,
xlabel=$\pi_{x^*y^*}$,
ylabel=$\langle \Hmin / \Ntot \rangle$]
\addplot[green!60!black, ultra thick, domain=0.25:0.95, samples=15] {0.244277707631205};
\addplot [color=green!60!black, mark=triangle*]
table[row sep=crcr]{
0.25	0\\
0.3	0\\
0.35	0\\
0.4	0\\
0.45	0\\
0.5	0\\
0.55	0\\
0.6	0\\
0.65	0\\
0.7	0\\
0.75	0\\
0.8	0\\
0.85	0.0164737659324713\\
0.9	0.068350237769307\\
0.95	0.094393721915263\\
};
\addplot [color=green!60!black, mark=*]
table[row sep=crcr]{
0.25	0\\
0.3	0\\
0.35	0\\
0.4	0\\
0.45	0\\
0.5	0\\
0.55	0\\
0.6	0\\
0.65	0\\
0.7	0\\
0.75	0\\
0.8	0\\
0.85	0.0164923172254225\\
0.9	0.0683374821930065\\
0.95	0.0941531119574111\\
};
\addplot[purple, domain=0.25:0.95, samples=15, double] {0.138997073548588};
\addplot [color=purple, mark=triangle, dashed, every mark/.append style={solid}]
table[row sep=crcr]{
0.25	0.133790102958313\\
0.3	0.133790102958313\\
0.35	0.133790102958313\\
0.4	0.133790102958313\\
0.45	0.133790102958313\\
0.5	0.133790102958313\\
0.55	0.133790102958313\\
0.6	0.133790102958313\\
0.65	0.133790102958313\\
0.7	0.133790102958313\\
0.75	0.133790102958313\\
0.8	0.133790102958313\\
0.85	0.133790102958313\\
0.9	0.133790102958313\\
0.95	0.133790102958313\\
};
\addplot[color=purple, mark=o, dashed, every mark/.append style={solid}]
table[row sep=crcr]{
0.25	0.133717960247309\\
0.3	0.133717960247309\\
0.35	0.133717960247309\\
0.4	0.133717960247309\\
0.45	0.133717960247309\\
0.5	0.133717960247309\\
0.55	0.133717960247309\\
0.6	0.133717960247309\\
0.65	0.133717960247309\\
0.7	0.133717960247309\\
0.75	0.133717960247309\\
0.8	0.133717960247309\\
0.85	0.133717960247309\\
0.9	0.133717960247309\\
0.95	0.133717960247309\\
};
\addplot [color=blue, densely dotted]
table[row sep=crcr]{
  0.25	0.129980786725109\\
0.3	0.129980786725109\\
0.35	0.129980786725109\\
0.4	0.129980786725109\\
0.45	0.129980786725109\\
0.5	0.129980786725109\\
0.55	0.129980786725109\\
0.6	0.129980786725109\\
0.65	0.129980786725109\\
0.7	0.129980786725109\\
0.75	0.129980786725109\\
0.8	0.129980786725109\\
0.85	0.129980786725109\\
0.9	0.129980786725109\\
0.95	0.129980786725109\\
};
\end{axis}
\end{tikzpicture}
\caption{$I_{\rm{CHSH}}=2.2587$}
\end{subfigure}
\begin{subfigure}{0.49\linewidth}
\begin{tikzpicture}
\begin{axis}[
/pgfplots/tick scale binop=\times,
/pgf/number format/.cd, 
fixed,
xmin=0.25,
xmax=0.95,
ymin=0,
ymax=0.85,
height=7cm,
xlabel=$\pi_{x^*y^*}$,
ylabel=$\langle \Hmin / \Ntot \rangle$]
\addplot[green!60!black, ultra thick, domain=0.25:0.95, samples=15] {0.809928739754258};
\addplot [color=green!60!black, mark=triangle*]
table[row sep=crcr]{%
0.25	0\\
0.3	0\\
0.35	0\\
0.4	0\\
0.45	0\\
0.5	0\\
0.55	9.1619418313739e-05\\
0.6	0.0489047270404283\\
0.65	0.116249331751384\\
0.7	0.179284424428422\\
0.75	0.236472461883836\\
0.8	0.282622454405125\\
0.85	0.307161313538826\\
0.9	0.284864792196845\\
0.95	0.091965648837619\\
};
\addplot [color=green!60!black, mark=*]
table[row sep=crcr]{%
0.25	0\\
0.3	0\\
0.35	0\\
0.4	0\\
0.45	0\\
0.5	0\\
0.55	4.68233151979241e-05\\
0.6	0.044206702145079\\
0.65	0.110466588518184\\
0.7	0.173026973562886\\
0.75	0.228576029743859\\
0.8	0.27383998292412\\
0.85	0.297939770438219\\
0.9	0.270576840614639\\
0.95	0.0748320058232833\\
};
\addplot[purple, domain=0.25:0.95, samples=15, double] {0.385530813537734};
\addplot [color=purple, mark=triangle, dashed, every mark/.append style={solid}]
table[row sep=crcr]{%
0.25	0.355916101654499\\
0.3	0.355916101654499\\
0.35	0.355916101654499\\
0.4	0.355916101654499\\
0.45	0.355916101654499\\
0.5	0.355916101654499\\
0.55	0.355916101654499\\
0.6	0.355916101654499\\
0.65	0.355916101654499\\
0.7	0.355916101654499\\
0.75	0.355916101654499\\
0.8	0.355916101654499\\
0.85	0.355916101654499\\
0.9	0.355916101654499\\
0.95	0.355916101654499\\
};
\addplot [color=purple, mark=o, dashed,every mark/.append style={solid}]
  table[row sep=crcr]{%
0.25	0.355002826537737\\
0.3	0.355002826537737\\
0.35	0.355002826537737\\
0.4	0.355002826537737\\
0.45	0.355002826537737\\
0.5	0.355002826537737\\
0.55	0.355002826537737\\
0.6	0.355002826537737\\
0.65	0.355002826537737\\
0.7	0.355002826537737\\
0.75	0.355002826537737\\
0.8	0.355002826537737\\
0.85	0.355002826537737\\
0.9	0.355002826537737\\
0.95	0.355002826537737\\
};
\addplot [color=blue, densely dotted]
table[row sep=crcr]{%
0.25	0.249460822796337\\
0.3	0.249460822796337\\
0.35	0.249460822796337\\
0.4	0.249460822796337\\
0.45	0.249460822796337\\
0.5	0.249460822796337\\
0.55	0.249460822796337\\
0.6	0.249460822796337\\
0.65	0.249460822796337\\
0.7	0.249460822796337\\
0.75	0.249460822796337\\
0.8	0.249460822796337\\
0.85	0.249460822796337\\
0.9	0.249460822796337\\
0.95	0.249460822796337\\
};
\end{axis}
\end{tikzpicture}
\caption{$I_{\rm{CHSH}}=2.4206$}
\end{subfigure}
\begin{subfigure}{0.49\linewidth}
\begin{tikzpicture}
\begin{axis}[
/pgfplots/tick scale binop=\times,
/pgf/number format/.cd, fixed,
xmin=0.25,
xmax=0.95,
ymin=0,
ymax=1.2,
height=7cm,
xlabel=$\pi_{x^*y^*}$,
ylabel=$\langle \Hmin / \Ntot \rangle$]
\addplot[green!60!black, ultra thick, domain=0.25:0.95, samples=15] {1.17058210790156};
\addplot [color=green!60!black, mark=triangle*]
  table[row sep=crcr]{
0.25	0\\
0.3	2.90787056517303e-05\\
0.35	0.0615746762762873\\
0.4	0.137941659295965\\
0.45	0.213379542843015\\
0.5	0.287977089143879\\
0.55	0.3618495286718\\
0.6	0.433829864630012\\
0.65	0.503383496225855\\
0.7	0.570235135973009\\
0.75	0.632321666280513\\
0.8	0.68619453014317\\
0.85	0.723491576581536\\
0.9	0.721172921585582\\
0.95	0.580902897581148\\
};
\addplot [color=green!60!black, mark=*]
table[row sep=crcr]{
0.25	0\\
0.3	1.85967161310366e-05\\
0.35	0.061645983440877\\
0.4	0.137998914687684\\
0.45	0.213096661707929\\
0.5	0.287659883554517\\
0.55	0.36184481328084\\
0.6	0.433596754448497\\
0.65	0.503844428188234\\
0.7	0.570881450375609\\
0.75	0.632229759001639\\
0.8	0.686230906936275\\
0.85	0.723225758894151\\
0.9	0.721562998278771\\
0.95	0.581264398949261\\
};
\addplot[purple, domain=0.25:0.95, samples=15, double] {0.516204956513184};
\addplot [color=purple, mark=triangle, dashed,  every mark/.append style={solid}]
table[row sep=crcr]{
0.25	0.495113500083617\\
0.3	0.495113500083617\\
0.35	0.495113500083617\\
0.4	0.495113500083617\\
0.45	0.495113500083617\\
0.5	0.495113500083617\\
0.55	0.495113500083617\\
0.6	0.495113500083617\\
0.65	0.495113500083617\\
0.7	0.495113500083617\\
0.75	0.495113500083617\\
0.8	0.495113500083617\\
0.85	0.495113500083617\\
0.9	0.495113500083617\\
0.95	0.495113500083617\\
};
\addplot[color=purple, mark=o,dashed,  every mark/.append style={solid}]
table[row sep=crcr]{
0.25	0.494429546326039\\
0.3	0.494429546326039\\
0.35	0.494429546326039\\
0.4	0.494429546326039\\
0.45	0.494429546326039\\
0.5	0.494429546326039\\
0.55	0.494429546326039\\
0.6	0.494429546326039\\
0.65	0.494429546326039\\
0.7	0.494429546326039\\
0.75	0.494429546326039\\
0.8	0.494429546326039\\
0.85	0.494429546326039\\
0.9	0.494429546326039\\
0.95	0.494429546326039\\
};
\addplot [color=blue, densely dotted]
table[row sep=crcr]{
0.25	0.452450699550399\\
0.3	0.452450699550399\\
0.35	0.452450699550399\\
0.4	0.452450699550399\\
0.45	0.452450699550399\\
0.5	0.452450699550399\\
0.55	0.452450699550399\\
0.6	0.452450699550399\\
0.65	0.452450699550399\\
0.7	0.452450699550399\\
0.75	0.452450699550399\\
0.8	0.452450699550399\\
0.85	0.452450699550399\\
0.9	0.452450699550399\\
0.95	0.452450699550399\\
};
\end{axis}
\end{tikzpicture}
\caption{$I_{\rm{CHSH}}=2.6090$}
\end{subfigure}
\caption{Average min-entropy rates as a function of the input distribution. In most cases, both regularisation methods give the same value for $\chi_{\textrm{one}}$, which is why they cannot be distinguished.}
\label{fig:inputbias}
\end{figure}
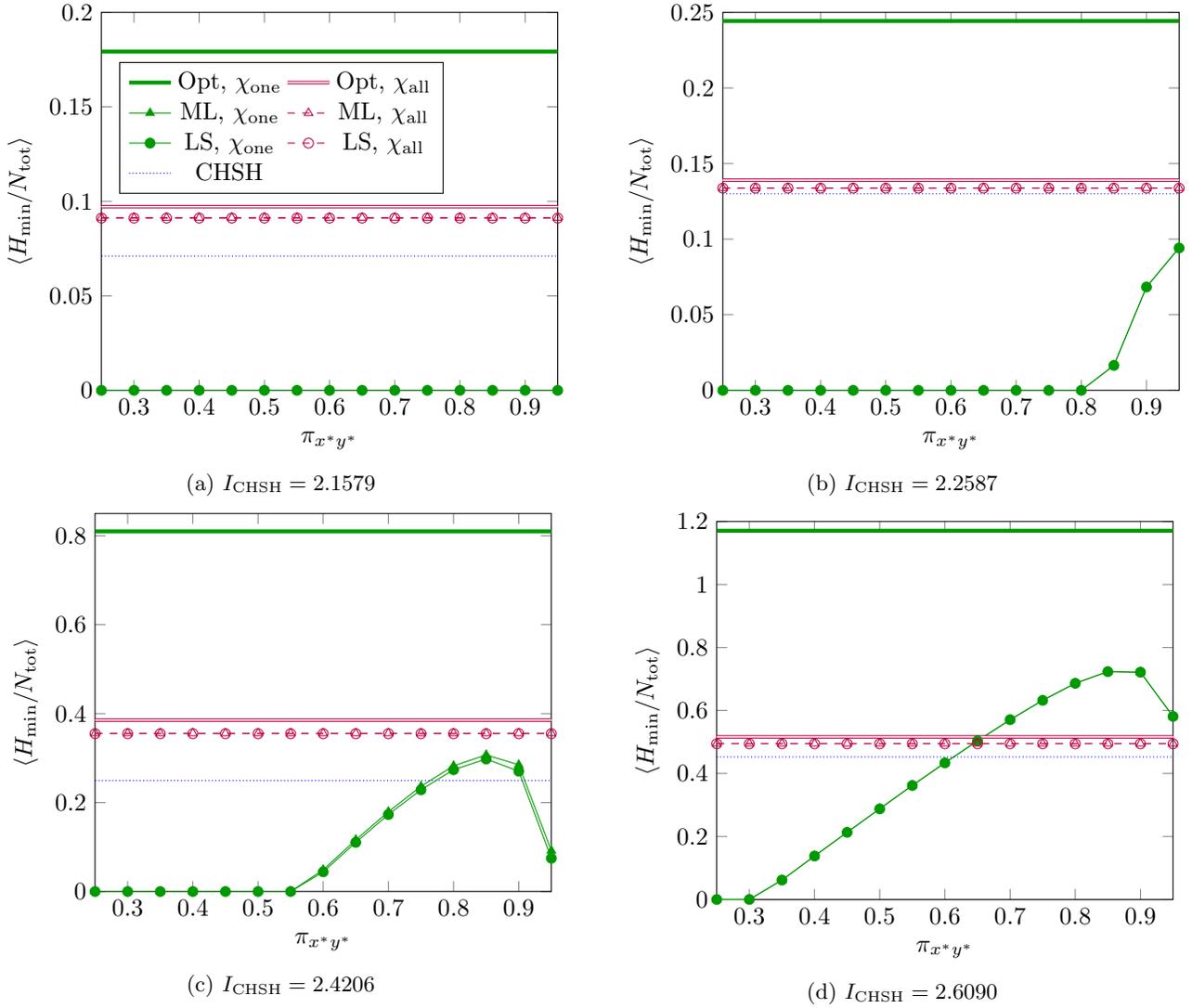

We observe that for three distributions, no advantage is obtained when generating randomness from only one input pair, independently of how the input distribution is biased towards that input pair. That confirms the observation based on the first graph: setting $\chi=\chi_{\textrm{one}}$ can give an advantage only for the behaviour with highest CHSH value. This is not surprising when one compares these results with the examples provided in~\cite{Nieto2018Device}, where the authors also observed that generating randomness from one input pair starts giving an advantage only for high enough $\Ntot > 10^8$. We thus decided not to use this possibility and to set $\chi = \chi_{\textrm{all}}$.

We then compared the min-entropy ratios obtained from the ML and LS regularisations. In that case, there is no varying parameter, so we decided to directly run the simulations described in Section~\ref{NumRes} for both regularisations, and to compare the obtained ratios $\langle \Hmin / \Hmin^{\rm{CHSH}} \rangle$. The results can be found in Figure \ref{fig:bothregularisations}. 

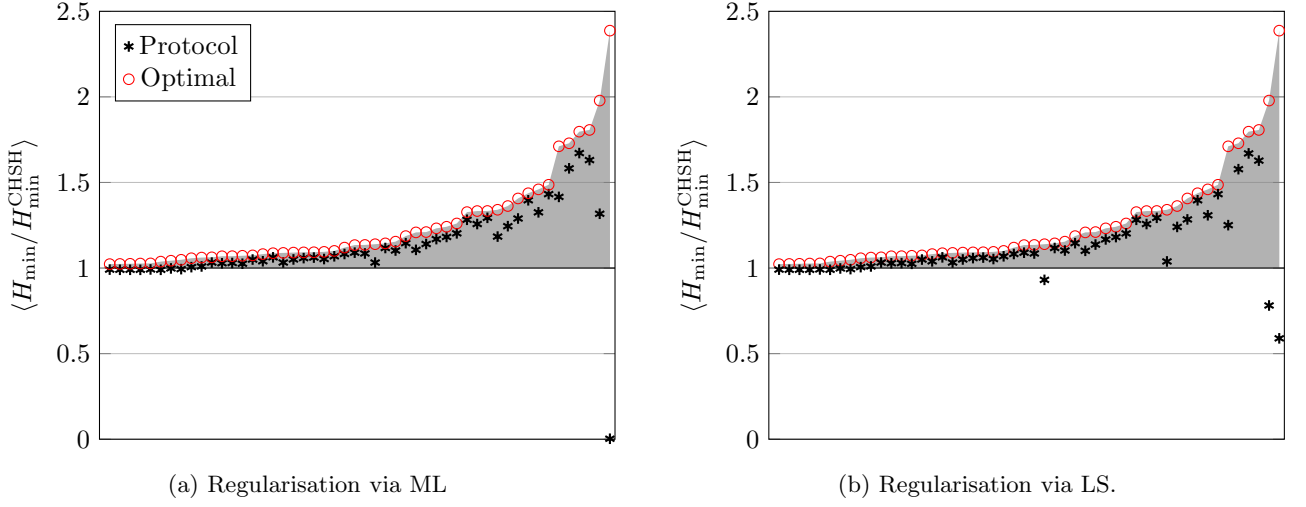
\begin{figure}
\begin{subfigure}{0.49\linewidth}
    \begin{tikzpicture}
      \begin{axis}[
        xmin=0,
        xmax=50.5,
        xtick=\empty,
        ymin=0,
        ymax=2.5,
        ylabel=$\langle \Hmin / \Hmin^{\rm{CHSH}} \rangle$, 
        ytick={0,0.5,1,1.5, 2,2.5},
        legend pos=north west, 
        grid=major
        ]
        \addlegendimage{only marks, mark=asterisk, black, thick}
        \addlegendimage{only marks, mark=o, red}
        \addplot [color=black, draw=none, thick, mark=asterisk]
        table[row sep=crcr]{
1	0.990848687291579\\
2	0.989840250679943\\
3	0.989868713404485\\
4	0.989672607343848\\
5	0.991828134714621\\
6	0.989687631845657\\
7	0.99775764186236\\
8	0.99369951076546\\
9	1.00534932003412\\
10	1.00931159716925\\
11	1.0319347775844\\
12	1.02791516870672\\
13	1.02936675708063\\
14	1.02527066043776\\
15	1.04967826513898\\
16	1.03967890409824\\
17	1.06254852909612\\
18	1.03226567312041\\
19	1.05030364274032\\
20	1.05811715817716\\
21	1.06121655294768\\
22	1.05264604385807\\
23	1.0687417501135\\
24	1.08217936422888\\
25	1.08988128232927\\
26	1.08508836755344\\
27	1.03209706386115\\
28	1.11713051411684\\
29	1.10236039282319\\
30	1.14538536422961\\
31	1.10629143855303\\
32	1.14053075597109\\
33	1.16911575041434\\
34	1.18166367922354\\
35	1.20273824177717\\
36	1.28195836436737\\
37	1.25715123005844\\
38	1.29342591267224\\
39	1.18384182510108\\
40	1.24461496883788\\
41	1.28948094044959\\
42	1.39587117689041\\
43	1.32552989273358\\
44	1.43182835015018\\
45	1.41570546591348\\
46	1.58261119954058\\
47	1.67172998409216\\
48	1.6306764773446\\
49	1.31714913141478\\
50	0.00279102803467605\\
        };
        \addlegendentry{Protocol};
        \addplot[name path=A, draw=none, color=red, mark=o]
        table[row sep=crcr]{
1	1.02301351215486\\
2	1.02396504069727\\
3	1.02495166337562\\
4	1.02637542156434\\
5	1.02717650406852\\
6	1.03756212959124\\
7	1.04289272579622\\
8	1.04831474747437\\
9	1.05728812433414\\
10	1.06122640329803\\
11	1.06267358464216\\
12	1.06894066712846\\
13	1.06940959556623\\
14	1.07178536016224\\
15	1.0739615420659\\
16	1.08154010793909\\
17	1.08675708170576\\
18	1.08855019979664\\
19	1.09033920071769\\
20	1.09162363541352\\
21	1.09237097025272\\
22	1.09460879526421\\
23	1.10018654291637\\
24	1.11946675195672\\
25	1.13421297600194\\
26	1.13505094197572\\
27	1.13926595808251\\
28	1.14416803603698\\
29	1.15425698912241\\
30	1.1866917176338\\
31	1.20826374375302\\
32	1.21012812172801\\
33	1.23186838518755\\
34	1.24126970760298\\
35	1.26043416989659\\
36	1.32685298948722\\
37	1.33299469320543\\
38	1.33341413236496\\
39	1.34050917695519\\
40	1.36224882216975\\
41	1.40653031129862\\
42	1.43703249246313\\
43	1.45961478647705\\
44	1.48653698065125\\
45	1.71112918136554\\
46	1.72891177035038\\
47	1.79672385876672\\
48	1.80676691022249\\
49	1.97841405945132\\
50	2.38689948288473\\
        };
        \addlegendentry{Optimal};
        \addplot [name path=B, black, domain={1:50.5}] {1};
        \addplot[gray, fill opacity=0.6]fill between[of=A and B, soft clip={domain=1:50}];
      \end{axis}
    \end{tikzpicture}
    \caption{Regularisation via ML}
\end{subfigure}
\begin{subfigure}{0.49\linewidth}
    \begin{tikzpicture}
      \begin{axis}[
        xmin=0,
        xmax=50.5,
        xtick=\empty,
        ymin=0,
        ymax=2.5,
        ylabel=$\langle \Hmin / \Hmin^{\rm{CHSH}} \rangle$, 
        ytick={0,0.5,1,1.5, 2,2.5},
        grid=major,
        legend pos=north west
        ]
        \addplot[name path=A, draw=none, color=red, mark=o]
        table[row sep=crcr]{
1	1.02301351215486\\
2	1.02396504069727\\
3	1.02495166337562\\
4	1.02637542156434\\
5	1.02717650406852\\
6	1.03756212959124\\
7	1.04289272579622\\
8	1.04831474747437\\
9	1.05728812433414\\
10	1.06122640329803\\
11	1.06267358464216\\
12	1.06894066712846\\
13	1.06940959556623\\
14	1.07178536016224\\
15	1.0739615420659\\
16	1.08154010793909\\
17	1.08675708170576\\
18	1.08855019979664\\
19	1.09033920071769\\
20	1.09162363541352\\
21	1.09237097025272\\
22	1.09460879526421\\
23	1.10018654291637\\
24	1.11946675195672\\
25	1.13421297600194\\
26	1.13505094197572\\
27	1.13926595808251\\
28	1.14416803603698\\
29	1.15425698912241\\
30	1.1866917176338\\
31	1.20826374375302\\
32	1.21012812172801\\
33	1.23186838518755\\
34	1.24126970760298\\
35	1.26043416989659\\
36	1.32685298948722\\
37	1.33299469320543\\
38	1.33341413236496\\
39	1.34050917695519\\
40	1.36224882216975\\
41	1.40653031129862\\
42	1.43703249246313\\
43	1.45961478647705\\
44	1.48653698065125\\
45	1.71112918136554\\
46	1.72891177035038\\
47	1.79672385876672\\
48	1.80676691022249\\
49	1.97841405945132\\
50	2.38689948288473\\
        };
        \addplot [name path=B, black, domain={1:50.5}] {1};
        \addplot[gray, fill opacity=0.6]fill between[of=A and B, soft clip={domain=1:50}];
        \addplot [color=black, draw=none, thick, mark=asterisk]
        table[row sep=crcr]{
1	0.990856495210405\\
2	0.989855414216932\\
3	0.98986871054268\\
4	0.989671832510987\\
5	0.991703541238189\\
6	0.989828366626902\\
7	0.997654686756302\\
8	0.993838407520263\\
9	1.00543402694283\\
10	1.00936067788039\\
11	1.03190984364254\\
12	1.02790251592579\\
13	1.02885390140545\\
14	1.02467957768128\\
15	1.04960755097172\\
16	1.03913081010971\\
17	1.06252165781859\\
18	1.03217282661767\\
19	1.05007619888424\\
20	1.05812005536058\\
21	1.06127472580482\\
22	1.05264894926284\\
23	1.06871994876682\\
24	1.08216112336045\\
25	1.08966617489539\\
26	1.08503156346768\\
27	0.930941580274364\\
28	1.117089762007\\
29	1.1024138197853\\
30	1.14548916038803\\
31	1.10073647088075\\
32	1.13719145315649\\
33	1.16719619507835\\
34	1.18096796340567\\
35	1.20232946923288\\
36	1.28183511938679\\
37	1.25718263052169\\
38	1.29344365305378\\
39	1.03860406610501\\
40	1.24028259704156\\
41	1.28410008709136\\
42	1.39583864758995\\
43	1.30836253446743\\
44	1.43171095327655\\
45	1.25062334270116\\
46	1.57760463758447\\
47	1.66964621926622\\
48	1.62754594304\\
49	0.781390824104349\\
50	0.589373544682185\\
        };
      \end{axis}
    \end{tikzpicture}
    \caption{Regularisation via LS.}
    \end{subfigure}
    \caption{Black asterisk: ratio between the rate obtained via our protocol and via the direct use of the CHSH inequality. Red circle: ratio between the maximal achievable min-entropy and the rate obtained via the direct use of the CHSH inequality.}
    \label{fig:bothregularisations}
\end{figure}

The ML regularisation performs better than the LS regularisation in $98\%$ of the cases. Moreover, while the protocol based on ML performs well for $98\%$ of the cases, that holds for LS only in $94\%$ of the cases. This leads us to claim that when one wants to regularise data in order to certify randomness, one should preferably minimise the KL divergence.  

\section{Generating randomness from one input pair}
\label{app:ChiOne}

To ensure that our method could result in better min-entropy bounds for $\chi=\chi_{\mathrm{one}}$ when the total number of rounds is big enough, we carried out the same simulations as the ones presented in the main text, but with $\Ntot=10^{12}$. In that case, our method allows us to identify which input pair $(x^*,y^*)$ yields the most favourable RB function, thanks to the ML regularised distribution. We then bias the input distribution towards that pair, setting $\pi_{x^*y^*}=0.99$. The results are presented in Figure~\ref{fig:OneInput}, where we plot the ratios between the min-entropy rate obtained via our protocol and via the direct use of the CHSH inequality $ \Hmin^{\rm{CHSH}}$, as well as the ratios between $-\log_2(G^{\chi}_{full}(P_{AB|XY}))$ and $\Hmin^{\rm{CHSH}}$, for $\chi=\chi_{\mathrm{one}}$ and $\chi=\chi_{\mathrm{all}}$. We highlighted in grey the region between these two ratios. $98\%$ of the simulations led to points falling in that region. In those cases, our protocol is good in two ways: not only it performs better than the direct use of CHSH, but it also achieves a higher ratio than the optimal one for all inputs. In that case, the advantage of our protocol is twofold: it allows us to identify the most favourable input pair, and then to tailor the Bell inequality to that specific input pair.  
 
\begin{figure}
 \begin{tikzpicture}
      \begin{axis}[
        xmin=0,
        xmax=50.5,
        xtick=\empty,
        ymin=0,
        ymax=8,
        ylabel=$\langle \Hmin / \Hmin^{\rm{CHSH}} \rangle$, 
        legend pos=north west, 
        grid=major
        ]
        \addlegendimage{only marks, mark=asterisk, black, thick}
        \addlegendimage{only marks, mark=o, red}
        \addlegendimage{only marks, mark=*, red, mark size=1pt}
        \addplot [color=black, draw=none, thick, mark=asterisk]
        table[row sep=crcr]{
1	1.13863680621994\\
2	1.07713112044323\\
3	1.09617911652522\\
4	1.05792834524014\\
5	1.199714163915\\
6	1.21124554258476\\
7	1.35930518314831\\
8	1.36303949104147\\
9	1.18370710256705\\
10	1.32388562529811\\
11	1.43080190307689\\
12	1.40166630658672\\
13	1.37563364494443\\
14	1.51967644486339\\
15	1.35995893849704\\
16	1.43414223859291\\
17	1.57528110418491\\
18	1.53299822798306\\
19	1.54247354216547\\
20	1.56395965699719\\
21	1.45832048659787\\
22	1.39299595797214\\
23	1.62922736487697\\
24	1.52296085996309\\
25	1.59936441442017\\
26	1.45156070387825\\
27	1.48089421854207\\
28	1.71184453668742\\
29	1.60193097435406\\
30	1.7481031187461\\
31	1.80649741744601\\
32	1.87327621983188\\
33	1.83136078584097\\
34	1.96328467823729\\
35	1.97812001899017\\
36	1.97699388291529\\
37	1.92573132021781\\
38	2.10495270234057\\
39	2.37426748367378\\
40	2.38470683929692\\
41	2.27641039893955\\
42	2.08492534709349\\
43	2.39740784820583\\
44	2.58669516583853\\
45	2.55831730584837\\
46	2.47781441674347\\
47	2.15667524675191\\
48	3.18491257216155\\
49	0.0865630217446855\\
50	1.95500122050745\\
};
        \addlegendentry{Protocol, $\chi_{\mathrm{one}}$};
        \addplot[name path=A, draw=none, color=red, mark=o]
        table[row sep=crcr]{
1	1.28174424902139\\
2	1.32665238823508\\
3	1.34066267923745\\
4	1.34612713009333\\
5	1.35300411842881\\
6	1.38036347429409\\
7	1.40462546339008\\
8	1.41864285350655\\
9	1.44129127041194\\
10	1.47605737717943\\
11	1.52073062947666\\
12	1.53539419884115\\
13	1.54896799920219\\
14	1.60557461511854\\
15	1.61892289066969\\
16	1.61950563458405\\
17	1.62827091501705\\
18	1.63502873923164\\
19	1.66563410275714\\
20	1.67720786532033\\
21	1.68320470473492\\
22	1.6852703709753\\
23	1.70026298232857\\
24	1.73043086112662\\
25	1.74201586604128\\
26	1.7623686375774\\
27	1.78356659110775\\
28	1.84593408520641\\
29	1.89303767942395\\
30	1.90125364423337\\
31	1.94366907354054\\
32	1.95894984310131\\
33	1.98492370085183\\
34	2.06696690218132\\
35	2.08764325634739\\
36	2.08894765831022\\
37	2.24191263623756\\
38	2.34878097130874\\
39	2.51685189245431\\
40	2.54475735255618\\
41	2.56754550569713\\
42	2.56860727022035\\
43	2.6235171926596\\
44	2.73810169883693\\
45	2.74549009486251\\
46	2.78435160966174\\
47	3.05750455215389\\
48	3.6287859687244\\
49	5.09411119993645\\
50	7.89966494451468\\
        };
        \addlegendentry{Optimal, $\chi_{\mathrm{one}}$};
        \addplot [name path=B,color=red, mark=*, draw=none, mark size = 1pt]
  table[row sep=crcr]{%
1	1.00453032663297\\
2	1.02374873720599\\
3	1.03129909504919\\
4	1.01106162045248\\
5	1.00084229673631\\
6	1.00179372335912\\
7	1.07530923833461\\
8	1.13233325971285\\
9	1.00324843165898\\
10	1.00593422081432\\
11	1.31435313026028\\
12	1.00497571487643\\
13	1.07136132944886\\
14	1.08520402443531\\
15	1.01376116878818\\
16	1.0442107874301\\
17	1.06359539839852\\
18	1.07502642808922\\
19	1.04545322398604\\
20	1.16848221125283\\
21	1.10093616992057\\
22	1.11764542680838\\
23	1.42159272317885\\
24	1.12355218567013\\
25	1.10055449182102\\
26	1.0470486993717\\
27	1.28534728675384\\
28	1.05035960231637\\
29	1.3160514038694\\
30	1.19379575655563\\
31	1.07149177518612\\
32	1.06829600080525\\
33	1.0513723015601\\
34	1.46988232700139\\
35	1.30848616800076\\
36	1.07701527023254\\
37	1.74999179198271\\
38	1.21090059243515\\
39	1.38671893696949\\
40	1.24147794625448\\
41	1.10633325429111\\
42	1.3243241392323\\
43	1.692601585014\\
44	1.1956665312095\\
45	1.44253489850395\\
46	1.78079958869001\\
47	1.64886526550826\\
48	1.21498483325549\\
49	2.32926976356033\\
50	1.90831333218319\\
};
\addlegendentry{Optimal, $\chi_{\mathrm{all}}$};
        \addplot [black, domain={1:50.5}] {1};
        \addplot[gray, fill opacity=0.6]fill between[of=A and B, soft clip={domain=1:50}];
      \end{axis}
    \end{tikzpicture}
\caption{Black asterisk: ratio between the rate obtained via our protocol for $\chi_{\mathrm{one}}$ and via the direct use of the CHSH inequality. Red circle: ratio between the maximal achievable min-entropy for $\chi_{\mathrm{one}}$ and the rate obtained via the direct use of the CHSH inequality. Red dot: ratio between the maximal achievable min-entropy for $\chi_{\mathrm{all}}$ and the rate obtained via the direct use of the CHSH inequality.}
\label{fig:OneInput}
\end{figure}
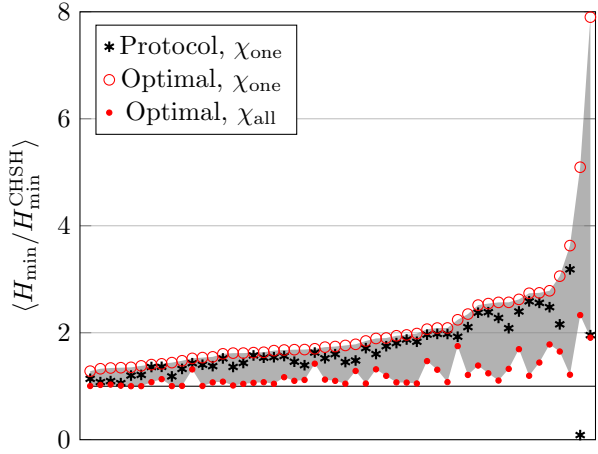

\end{widetext}
\end{document}